\documentclass[a4paper,onecolumn,11pt,accepted=2025-01-08]{quantumarticle}
\pdfoutput=1

\usepackage[utf8]{inputenc}
\usepackage[english]{babel}
\usepackage[T1]{fontenc}
\usepackage[sorting=none,
date=year,
backend=bibtex,
style=numeric-comp,
giveninits=true,
url=false,
doi=false,
maxnames=10,
isbn=false]{biblatex}
\usepackage{amssymb,amsmath,amsfonts,amsthm,wasysym}
\usepackage{amsbsy} 
\usepackage{epsfig}
\usepackage{latexsym}
\usepackage{comment}
\usepackage{tensor}
\usepackage{adjustbox}
\usepackage{graphicx}
\usepackage[]{cancel}
\usepackage{xcolor}
\usepackage{enumitem}
\usepackage{subcaption}
\usepackage{hyperref}
\usepackage{booktabs}
\usepackage{mathtools}
\usepackage{csquotes}

\def\be{\begin{equation}}
	\def\ee{\end{equation}}
\def\dd{{\rm d}}
\def\bes{\begin{eqnarray}}
	\def\ees{\end{eqnarray}}
	
\DeclareMathOperator{\sgn}{sgn}

\newcommand{\intsum}{\,\,\,\mathclap{\displaystyle\int}\mathclap{\textstyle\sum}}

\DeclareFieldFormat{eprint:arxiv}{%
	arXiv\addcolon\space
	\ifhyperref
	{\href{https://arxiv.org/abs/#1}{\texttt{#1}%
			\iffieldundef{eprintclass}%
			{}%
			{ \texttt{\mkbibbrackets{\thefield{eprintclass}}}\kern-0.1em}}}%
	{\texttt{#1}%
		\iffieldundef{eprintclass}%
		{}%
		{ \texttt{\mkbibbrackets{\thefield{eprintclass}}}\kern-0.1em}}}%

\DeclareBibliographyDriver{article}{%
	\printnames{author}%
	\setunit{\labelnamepunct}\newblock
	\printfield{title}%
	\iffieldundef{journaltitle}
	{\setunit{\addspace}}
	{%
		\newunit\newblock
		\printfield{journaltitle}%
		\setunit{\space}%
		\iffieldundef{volume}
		{}
		{%
			\printfield{volume}%
			\setunit{\addcomma\space}%
		}%
		\iffieldundef{pages}
		{}
		{%
			\printfield{pages}%
			\setunit{\space}%
		}%
	}%
	\printtext[parens]{\printfield{year}}%
	\setunit{\addcomma\space}
	\iffieldundef{eprinttype}
	{}
	{\printfield[eprint:\strfield{eprinttype}]{eprint}}
	\newunit\newblock
	\printfield{note}%
	\finentry}

\DeclareFieldFormat[article]{volume}{\textbf{#1}}

\DeclareBibliographyDriver{book}{%
	\printnames{author}%
	\setunit{\labelnamepunct}\newblock
	\printfield{title}%
	\iffieldundef{series}
	{} %
	{\setunit{\addcomma\space}\printfield{series}} % 
	\setunit{\addspace}
	\printtext[parens]{%
		\printlist{publisher}%
		\setunit{\addcomma\space}
		\printlist{location}%
		\setunit{\addcomma\space}
		\printfield{year}}%
	\finentry}

\DeclareFieldFormat{title}{%
	\iffieldundef{doi}{\emph{#1}}{%
		\href{https://doi.org/\thefield{doi}}{\emph{#1}}%
	}%
}

\DeclareBibliographyDriver{inproceedings}{%
	\printnames{author}%
	\setunit{\labelnamepunct}\newblock
	\printfield{title}%
	\setunit{\addcomma\space}  %
	\printfield{booktitle}%
	\setunit{\addcomma\space}
	\printtext{pp.\space}%
	\printfield{pages}%
	\newunit\newblock
	\printtext{Ed. by\space}%
	\printnames[editorincollection]{editor}%
	\iffieldundef{series}
	{}
	{\printfield{series}%
		\setunit{\addcomma\space}}
	\iffieldundef{volume}
	{}
	{\printfield{volume}%
		\setunit{\addcomma\space}}
	\iffieldundef{number}
	{}
	{\printfield{number}%
		\setunit{\addcomma\space}}
	\setunit{\addspace}%
	\printtext[parens]{%
		\printlist{publisher}%
		\setunit{\addcomma\space}
		\printlist{location}%
		\setunit{\addcomma\space}
		\printfield{year}}%
	\iffieldundef{eprinttype}
	{}
	{\setunit{\addcomma\space}
		\printfield[eprint:\strfield{eprinttype}]{eprint}}
	\newunit\newblock
	\printfield{note}
	\finentry}

\DeclareFieldFormat{booktitle}{%
	\iffieldundef{doi}{\emph{#1}}{%
		\href{https://doi.org/\thefield{doi}}{\emph{#1}}%
	}%
}

\renewbibmacro{in:}{}

\DeclareFieldFormat{journaltitle}{%
	\iffieldundef{doi}{\emph{#1}}{%
		\href{https://doi.org/\thefield{doi}}{\emph{#1}}%
	}%
}

\DefineBibliographyStrings{english}{%
	page             = {},
	pages            = {},
} 

\AtEveryBibitem{\clearfield{number}}
\AtEveryBibitem{\clearfield{issue}}

\begin{document}
	
\title{Relational dynamics and Page--Wootters formalism in group field theory}
	
\author{Andrea Calcinari}
\affiliation{School of Mathematical and Physical Sciences, University of Sheffield, Hicks Building, Hounsfield Road, Sheffield S3 7RH, United Kingdom}
\affiliation{Departamento de Física Teórica and IPARCOS, Facultad de Ciencias Físicas, Universidad Complutense de Madrid, Plaza de las Ciencias 1, Madrid 28040, Spain}
\email{andrcalc@ucm.es}
\orcid{0000-0003-3028-0587}

\author{Steffen Gielen}
\email{s.c.gielen@sheffield.ac.uk}
\affiliation{School of Mathematical and Physical Sciences, University of Sheffield, Hicks Building, Hounsfield Road, Sheffield S3 7RH, United Kingdom}
\orcid{0000-0002-8653-5430}

	\maketitle
	
	\begin{abstract}
Group field theory posits that spacetime is emergent and is hence defined without any background notion of space or time; dynamical questions are formulated in relational terms, in particular using (scalar) matter degrees of freedom as time. Unlike in canonical quantisation of gravitational systems, there is no obvious notion of coordinate transformations or constraints, and established quantisation methods cannot be directly applied. As a result, different canonical formalisms for group field theory have been discussed in the literature. We address these issues using a parametrised version of group field theory, in which all (geometry and matter) degrees of freedom evolve in a fiducial parameter. There is a constraint associated to the freedom of reparametrisation and
the Dirac quantisation programme can be implemented. Using the ``trinity of relational dynamics'', we show that the resulting ``clock-neutral'' theory is entirely equivalent to a deparametrised canonical group field theory, interpreted in terms of the Page--Wootters formalism. Our results not only show that the deparametrised quantisation is fully covariant and can be seen as encoding the dynamics of joint quantum matter and geometry degrees of freedom, they also appear to be the first application of the Page--Wootters formalism directly to non-perturbative quantum gravity. We show extensions to a setting in which many independent gauge symmetries are introduced, which connects to the ``multi-fingered time'' idea in quantum gravity and provides a somewhat novel extension of the Page--Wootters formalism.
	\end{abstract}

\section{Introduction}

A central theme in many approaches to quantum gravity is that of background independence. This principle stems directly from general relativity where the geometry of spacetime is not taken as a background structure for a given system, but rather is understood as a dynamical part of it. In particular, the absence of a background time poses a challenge for the definition of dynamics, which leads to the ``problem of time'' of classical and quantum gravity \cite{Isham:1992ms,Kuchar:1991qf}. More precisely, the canonical Hamiltonian of general relativity vanishes on shell and gravitational observables, required to have vanishing Poisson brackets with the constraints, appear to be ``frozen'' in coordinate time \cite{Arnowitt:1962hi,Diracbook,Gaugebook}. The most common proposal to bypass this problem is to adopt a relational strategy, where one picks a degree of freedom of the system to serve as internal time relative to which the remaining degrees of freedom evolve. For example, in the case of cosmological settings, the proposal is usually to add matter (commonly a massless scalar field) to describe the relational dynamics of the gravitational degrees of freedom. At the classical level, one can define relational Dirac observables of constrained systems following the general theory of \cite{DittrichDO,GieselDO,Tambornino} (see also \cite{Rovelli_Amodel,Rovelli_Anhypothesis,RovelliOBS,RovelliPO} for earlier work and \cite{Goeller:2022rsx} for a recent extension of the concept of relational observables), which allows to deal with the freedom of choosing different clocks by means of ``complete observables''; these implement precisely the idea of encoding dynamics as a relation between phase space functions, without referring to any external structure (in particular any background time). The quantum theory for constrained systems can be obtained following the Dirac quantisation programme \cite{Matschull:1996up,Diracbook,Gaugebook}, which has the advantage of preserving the structures of the classical theory such as constraints (which, in the case of gravity, are associated with the notion of covariance). The Dirac programme provides a clear notion of relational observables for the quantum theory and importantly, because all the degrees of freedom are treated on the same footing, does not require any gauge fixing or choice of time parameter before quantisation. This aspect is clarified in particular in the recent work of \cite{Trinity,RelativisticTrinity}, where the Dirac quantisation programme is denoted ``clock-neutral'' as it describes physics before choosing a (temporal) reference frame. 

A prominent example of a background-independent framework for quantum gravity is loop quantum gravity (LQG) \cite{Ashtekar:2021kfp}. In its canonical formulation, LQG seeks to obtain a quantisation of general relativity by means of the Dirac programme \cite{ThiemannBook}: one formally defines a space of physical quantum states by imposing constraints on kinematical states. Although the full realisation of this programme faces significant technical challenges, notable progress has been made in the context of loop quantum cosmology \cite{LQCreview} where, due to symmetry reduction, there is only one constraint associated with the freedom to choose the time parameter. Another approach that adheres to the paradigm of background independence is given by group field theory (GFT), which describes spacetime as emerging from the collective behaviour of (possibly pre-geometric) quantum gravity degrees of freedom \cite{FreidelGFT,Oriti:2011jm}. While GFT is closely related to the covariant spin foam formulation of LQG \cite{Reisenberger_2000,PerezSFQG} and to tensor models \cite{DiFrancesco:1993cyw,ColourTensor}, a canonical framework was also proposed in \cite{Oriti_GFT2ndLQG}. This Hilbert space quantisation of GFT is motivated by similarities with the canonical setting of LQG and, more practically, is used to extract effective cosmological dynamics \cite{GFTcosmoLONGpaper,Gielen_2016}.

A somewhat peculiar challenge for the GFT framework is that one cannot directly apply the methods of canonical quantisation, due to the absence of a Hamiltonian formulation of the theory. Since there is no background time and no immediate definition of a phase space structure at the classical level, the Hilbert space formalism for GFT of \cite{Oriti_GFT2ndLQG} is not derived from a canonical quantisation of a classical theory; rather, it is introduced via the kinematical structures of a Fock space, constructed along the lines of many-body quantum physics. This is the so-called \textit{algebraic approach} to GFT, a canonical formulation where the equations of motion are imposed at the quantum level as constraints, so as to reduce from the postulated kinematical Hilbert space to a physical Hilbert space. Another way to define dynamics in GFT is to follow a \textit{deparametrised approach} \cite{relham_Wilson_Ewing_2019,relhamadd}: this amounts to selecting a time parameter for the classical theory which allows to write down a relational Hamiltonian, and hence perform a standard canonical quantisation. Deparametrisation is subject to the general concerns and criticisms of ``tempus ante quantum'' frameworks \cite{Isham:1992ms,Kuchar:1991qf}, as it is not clear whether the choice of a classical time label before quantisation breaks clock-covariance \cite{Marchetti2021}. While both of these frameworks yield similar dynamics in the restriction to homogeneous and isotropic cosmology (see, e.g., \cite{Gielen_2020}), one would ideally like to leverage their complementary strengths by performing a genuine canonical quantisation of a classical theory in terms of constraints, without singling out an arbitrary time parameter at the classical level.

In this paper we address these challenges by defining relational dynamics in GFT, in particular for models where a scalar field is used as relational clock, in a way that connects with the known methods for quantisation (such as the Dirac programme mentioned above and the Page--Wootters formalism, as explained below). We will only be interested in the free theory, where interactions are ignored. Following the parametrisation strategy adopted in quantum mechanics \cite{Diracbook,Gaugebook} and quantum field theories \cite{Kuchar:1989bk,Kuchar:1989wz,Varadarajan:2006am}, we reformulate GFT cosmology models as constrained systems where the constraint is associated with the notion of time reparametrisation invariance. This allows us to implement the programme of Dirac quantisation along the lines of other systems that are well understood (e.g., loop quantum cosmology), and to connect with the ``trinity of relational quantum dynamics'' of \cite{Trinity,RelativisticTrinity}. Already for the classical theory, we can define relational Dirac observables for GFT in a precise way as those that Poisson-commute with the constraint \cite{DittrichDO,GieselDO,Tambornino}. Quantising the parametrised GFT à la Dirac, namely reducing from a kinematical Hilbert space to a physical space via group averaging techniques, makes clock covariance transparent for the GFT cosmological models of interest. In particular, we define the relational Dirac observable associated with the GFT number operator (the main observable for cosmology), and interpret its quantum dynamics by means of the Page--Wootters formalism \cite{PW,Wootters}. More precisely, thanks to the equivalence established in \cite{Trinity,RelativisticTrinity}, we describe the expectation value of the GFT number operator as conditional on the reading of the (quantum) clock associated with the matter scalar field. This is the first application of the Page--Wootters formalism in non-perturbative quantum gravity (a recent application to a perturbative quantum gravity setting was given in \cite{DeVuyst:2024pop}), and enables us to discuss GFT cosmology in a ``tempus post quantum'' framework \cite{Isham:1992ms,Kuchar:1991qf,Marchetti2021}. Remarkably, the relational dynamics turn out to match with the ones obtained in the deparametrised setting (where the clock is selected prior to quantisation), proving that deparametrisation in GFT cosmology is fully covariant. By defining a new variant of the Page--Wootters formalism for the case of multiple quantum clocks, we also generalise the setup to a situation with multiple Hamiltonian constraints (associated to an independent gauge invariance for each field mode), which realises the idea of ``multi-fingered time evolution'' \cite{Kuchar_Bubble,Kuchar_generic}.

Ultimately, our results establish a framework that consistently describes the relational evolution of GFT geometric observables with respect to a ``quantum time'', here identified with the matter scalar field. Crucially, this is given by a canonical quantisation that does not require to single out the clock classically, and where dynamics are implemented by a Hamiltonian constraint (somewhat similar to the situation in general relativity). We thus obtain a manifestly covariant formulation of GFT relational quantum dynamics, which in particular is equipped with the conditional interpretation of the Page--Wootters formalism, providing robust insights on relational dynamics for quantum gravity.

The paper is structured as follows. In section \ref{GFTSec} we briefly review the main approaches adopted in the literature to define relational dynamics in GFT cosmology, emphasising relative merits and limitations. In section \ref{ClassicalSec} we parametrise the GFT models of interest, focussing on a single field mode for simplicity, and we discuss the notion of classical Dirac observables. We then quantise the theory in section \ref{QuantumSec}, clearly distinguishing between kinematical aspects and relational quantum dynamics, obtained equivalently with the Dirac and the Page--Wootters formalisms. The construction is generalised to the case of multiple field modes in section \ref{MultimodeSec}, where we showcase two scenarios: one where all the modes evolve with respect to one clock, and one where they evolve with respect to separate ``single-mode times''. Appendix \ref{AppA} shows details on the relational GFT Hamiltonian introduced in \cite{relham_Wilson_Ewing_2019,relhamadd} and its full spectral decomposition.

\section{Relational dynamics in group field theory}\label{GFTSec}

The GFT formalism was originally conceived as a way of encoding discrete gravity (spin foam) amplitudes into Feynman amplitudes of an abstract field theory on a group manifold \cite{FreidelGFT,Oriti:2011jm}. The quantum theory is then defined as a path integral or partition function, defined perturbatively in terms of Feynman diagrams or perhaps non-perturbatively using resummation methods analogous to matrix and tensor models \cite{DiFrancesco:1993cyw,ColourTensor}.

Concretely, models for simplicial quantum gravity coupled to a scalar field $\chi$ can be built by defining a group field $\varphi$ whose arguments are four ${\rm SU}(2)$ elements and a real label $\chi$, satisfying
\be
\varphi(g_I,\chi) \equiv \varphi(g_1,\dots,g_4,\chi) = \varphi (g_1 h, \dots, g_4 h, \chi)\quad\forall \;h \in SU(2)\,.
\ee
This property ensures that the resulting amplitudes are invariant under local ${\rm SU}(2)$ gauge transformations acting on vertices. One often prefers working with field modes obtained from the Peter--Weyl decomposition
\be\label{peterw}
\varphi(g_I,\chi) = \sum_J \varphi_J(\chi) D_J(g_I)\,, \quad 	D_J(g_I) = \sum_{n_I} R_{n_I}^{j_I, \imath} \prod_{a=1}^{4} \sqrt{2j_a+1} D^{(j_a)}_{m_a, n_a}(g_a)\,,
\ee
where $J=(j_I, m_I, \imath )$ is a multi-index containing irreducible ${\rm SU}(2)$ representations $j_I$, magnetic indices $m_I$ and intertwiner labels $\imath$ labelling the intertwiners $R_{n_I}^{j_I, \imath}$; $D^{(j)}_{m, n}(g)$ are Wigner $D$-matrices in the $j$ representation. This choice of kinematical structure ensures that the resulting Feynman amplitudes will contain the same variables as the simplicial gravity model of interest; the action can be chosen so that the amplitudes agree \cite{Reisenberger_2000}.

More recently, significant attention has been focused on Hilbert space formalisms for GFT, starting from the proposal of \cite{Oriti_GFT2ndLQG} which was immediately applied to obtain effective GFT cosmology \cite{GFTcosmoLONGpaper,Gielen_2016}. As in the somewhat related setting of loop quantum gravity and its application to cosmology \cite{LQCreview}, a Hilbert space formalism has the advantage of giving easier access to dynamical equations which can be related to the equations of cosmology. Since GFT is a background-independent approach, one would expect any such Hilbert space definition of the theory to suffer from the problem of time. The main proposal in GFT has been to follow what is done in other approaches to quantum gravity and use matter fields as relational clocks; here this role is played by the scalar field variable $\chi$. The question is then how precisely to obtain relational dynamics in GFT, and two main approaches have been followed.

{\bf Algebraic approach.} In this approach one uses a complex group field, and assumes that the field operators $\hat{\varphi}$ and $\hat{\varphi}^\dagger$ satisfy the commutation relation
\begin{equation}
	\left[\hat{\varphi}_J (\chi) , \hat{\varphi}_{J'}^\dagger(\chi')\right] = \delta_{JJ'}\delta(\chi-\chi')\,.
\end{equation}
These operators act as creation and annihilation operators to generate a (kinematical) Fock space, based on a vacuum $|\emptyset\rangle$ that satisfies $\hat{\varphi}_J(\chi)|\emptyset\rangle = 0$ for all $J$ and $\chi$; a one-particle state reads $|J,\chi\rangle = \hat{\varphi}^\dagger_J(\chi)|\emptyset\rangle$. By analogy with spin network states of loop quantum gravity, the ``particle'' is now interpreted as a tetrahedron or open four-valent spin network node, and more complicated spin network states can be constructed from (not uniquely defined) many-particle states \cite{Oriti_GFT2ndLQG}. Hence, these Fock states can be related to kinematical states in loop quantum gravity \cite{Oriti_GFTandLQG}.

One can construct operators on such a Hilbert space, such as the number operator
\begin{equation}\label{N}
	\hat{N} = \sum_J \int \dd \chi \, \hat{\varphi}^\dagger_J(\chi ) \hat{\varphi}_J(\chi) = \sum_J \int \dd \chi \,  \hat{N}_J(\chi)\,,
\end{equation}
and the volume operator $\hat{V} = \sum_J v_J  \int \dd \chi \, \hat{N}_J(\chi)$ where $v_J$ is the volume assigned to quantum tetrahedra with group theoretic data $J$ (usually imported from loop quantum gravity). Furthermore, one can define a scalar field operator and scalar field momentum \cite{Oriti_2016,BOriti_2017,Marchetti2021}
\be \label{XP}
\hat{X} = \sum_J \int \dd \chi \, \chi \, \hat{\varphi}_J^\dagger(\chi) \,  \hat{\varphi}_J(\chi)   \,, \qquad \hat{\Pi}  = -{\rm i} \sum_J \int \dd \chi \, \left( \hat{\varphi}^\dagger_J(\chi) \partial_\chi \hat{\varphi}_J(\chi)   \right) \,,
\ee
which satisfy $ [\hat{X},\hat{\Pi}] = {\rm i} \hat{N} $, meaning that these operators are not exactly canonically conjugate. One can indeed make the observation that both operators are ``extensive''  whereas one would expect one intensive and one extensive quantity to form a canonical pair (see \cite{Gielen:2014uga} for a related discussion).

These operators act on the ``frozen'' kinematical Hilbert space and do not encode any notion of dynamics. The proposal of \cite{Oriti_2016,BOriti_2017} was to define relational observables by removing the $\chi$ integral from the definition of basic operators; one obtains, e.g., a relational number operator by
\be
\hat{N}(\chi) =\sum_J \hat{\varphi}^\dagger_J(\chi ) \hat{\varphi}_J(\chi) \,.
\ee
Expectation values of such an operator in mean-field coherent states appear well-behaved and follow the behaviour of classical solutions, leading to effective Friedmann equations for a relational volume operator $\hat{V}(\chi)$ which provide excellent agreement with usual cosmology \cite{Oriti_2016,BOriti_2017}. However, the distributional nature of such an operator becomes clear once one studies higher powers and divergences appear \cite{Isha_thermalGFT,Isha_thermalGFT2}; for less simple states already expectation values are divergent \cite{Gauss}.

The usual assumption is that physical states should satisfy constraints of the form 
\begin{equation}
	\frac{\delta S[\hat{\varphi},\hat{\varphi}^\dagger]}{\delta \hat{\varphi}^\dagger_J(\chi)}| \Psi \rangle =0
	\label{constraint}
\end{equation}
corresponding to the quantum equations of motion of the theory \cite{Gielen_2016,GFTcosmoLONGpaper,Oriti_2016,BOriti_2017}. This looks similar to the type of equation used in a Dirac quantisation of constrained systems. For a simple choice of action, one can find exact solutions to this constraint equation \cite{Gauss} (see also \cite{GFTcosmoLONGpaper} for an exact solution of an interacting theory). Such states will be non-normalisable in the original Fock space inner product, leading to another type of divergence.\footnote{Moreover, the equation of motion and its Hermitian conjugate are expected to be second class constraints with no joint solutions.}

One proposal for dealing with these issues with divergences and obtaining a different notion of relational dynamics in the algebraic approach is to focus on very specific types of states, called coherent peaked states \cite{Marchetti2021}. These states are defined as coherent states
\be
|\sigma_\epsilon ; \chi_0, \pi_0\rangle = \mathcal{N}_{{\sigma}} \exp\left(\sum_J \int {\rm d} \chi \, {\sigma}^\epsilon_J (\chi) \hat{\varphi}_J^\dagger(\chi) \right)|\emptyset\rangle \,, \qquad \mathcal{N}_{{\sigma}} = \exp\left( -\frac{1}{2} \sum_J \int {\rm d} \chi |{\sigma}^\epsilon_J(\chi)|^{2} \right) \,,
\ee
with mean field
\be
\sigma^\epsilon_J (\chi) = e^{-\frac{(\chi-\chi_0)^2}{2\epsilon}} e^{{\rm i} \pi_0 (\chi-\chi_0)}\tilde{\sigma}_J(\chi)
\ee
defined as the product of a Gaussian and a function $\tilde{\sigma}_J(\chi)$ assumed to satisfy certain equations (viewed as approximations to the full equations of motion) so that the state is now normalisable in the Fock space inner product. 

The next step is then to compute expectation values of operators such as (\ref{N}) and (\ref{XP}) and view them as functions of the parameter $\chi_0$ used in the definition of the state. Concretely, given that in such a peaked coherent state $\langle \hat{X}\rangle \simeq \chi_0 \langle\hat{N}\rangle$, one now defines an intensive effective clock variable by
\be\label{redefchi}
\hat{\chi} = \frac{\hat{X}}{\langle\hat{N}\rangle} 
\ee
so that in the given approximations $\langle\hat{\chi}\rangle = \chi_0$, and $\chi_0$ can be seen as the expectation value of the clock $\hat\chi$.  One can also show that  $\langle \hat\Pi \rangle \simeq \langle \hat{H}\rangle $ 
for a suitably defined $\hat{H}$ (again adapted to the specific choice of state) and finally obtain an effective Heisenberg-like equation for the volume operator,
\begin{equation}
	\frac{\dd }{\dd \chi_0}  \langle \hat{V}\rangle =  {\rm i}\langle[\hat{H} , \hat{V} ] \rangle\,.
\end{equation} 
In this picture, dynamical equations emerge on the kinematical Hilbert space from considering the evolution of expectation values relative to the clock expectation value $\langle\hat{\chi}\rangle$, choosing rather specific states so that a simple relation between these expectation values can be derived. The proposal does not refer to a classical Hamiltonian notion of dynamics or to classical relational observables as in usual Dirac quantisation, and it is unclear whether constraints such as (\ref{constraint}) play a significant role. The resulting equations for relational dynamics rely on a particular choice of state.

\textbf{Deparametrised approach.} In contrast to the algebraic approach, the deparametrised approach is a conventional canonical quantisation of a GFT action, using the scalar field label $\chi$ as a notion of time variable from the beginning. This approach was first developed in \cite{relham_Wilson_Ewing_2019,relhamadd}, based on a real group field whose Peter--Weyl modes satisfy $\overline{\varphi_J(\chi)} = (-1)^{\sum_I(j_I-m_I)}\varphi_{-J}(\chi)$ with $-J=(j_I,- m_I, \imath )$ denoting sign reversal of magnetic indices.

In the most commonly used situation, one restricts the action to its quadratic part, assuming interactions are weak or entirely negligible. One also assumes that the kinetic term is made up of a ``mass term'' with no derivatives and a term with second derivatives in $\chi$. First derivatives are excluded if one imposes symmetry under $\chi\rightarrow -\chi$, and explicit $\chi$ dependence is forbidden by a shift symmetry $\chi\rightarrow\chi+\chi_0$. It is natural to demand these symmetries given that $\chi$ should represent a free massless scalar field, which has the same symmetries, in GFT \cite{Oriti_2016,BOriti_2017}. The action then reads
\be
S_0[\varphi] = \frac{1}{2}\int \dd \chi \sum_J \varphi_{-J} (\chi) \left(K_J^{(0)} + K_J^{(2)} \partial^2_\chi \right) \varphi_J(\chi) \,,	
\ee
where $K_J^{(0)}$ and $K_J^{(2)}$ can in general be positive or negative. In the following we take the kinetic term to be symmetric under $J\leftrightarrow-J$, i.e., $K_J^{(0)}=K_{-J}^{(0)}$ and $K_J^{(2)}=K_{-J}^{(2)}$ . This kinetic term is also the one appearing in studies of GFT phase transitions \cite{Marchetti:2022igl,Marchetti:2022nrf}. After integration by parts, one has
\be\label{DepAction}
S_0[\varphi] = \frac{1}{2}\int \dd \chi \sum_J \left(K_J^{(0)}\varphi_{-J}(\chi)\varphi_J(\chi)-  K_J^{(2)}\partial_\chi\varphi_{-J}(\chi)\partial_\chi\varphi_J(\chi) \right) \,,
\ee
which is now just a function of field modes and their ``time'' derivatives, so that the Legendre transform to a relational Hamiltonian is straightforward. Introducing the conjugate momentum $\pi_J(\chi)$, one obtains
\begin{equation}\label{relham}
	{H} = -\frac{1}{2} \sum_{J} \left[\frac{\pi_{J}(\chi)\pi_{-J}(\chi)}{K^{(2)}_{J}}+K^{(0)}_{J}\varphi_{J}(\chi)\varphi_{-J}(\chi)\right] \,.
\end{equation}
$\chi$ appears on the same footing as a background time parameter, and the equation of motion of any classical observable $\mathcal{O}$ can be derived from the Poisson bracket ${\dd \mathcal{O}}/{\dd \chi} =	\{ \mathcal{O} , {H} \}$. Only now, the group field and its momentum are promoted to operators with the canonical equal-time commutation relation
\begin{equation}
	\left[ \hat{\varphi}_J(\chi) , \hat{\pi}_{J'}(\chi)  \right] = {\rm i} \delta_{JJ'} \,.
\end{equation}
All operators satisfy the Heisenberg equations of motion
\begin{equation}\label{Heisen}
	{\rm i} \frac{\dd \hat{\mathcal{O}}}{\dd \chi} = [\hat{\mathcal{O}} , \hat{{H}}] \,,
\end{equation}
where $\hat{{H}}$ is the quantum version of \eqref{relham}. 

Depending on relative signs of $K_J^{(0)}$ and $K_J^{(2)}$, the Hamiltonian for each $J$ mode is either a harmonic oscillator or an upside-down harmonic oscillator, which after introducing the usual creation and annihilation operators takes the form of a squeezing operator \cite{Gielen_2020}. Concretely, one defines
\begin{equation}\label{aadag}
	\hat{a}_J = \frac{1}{\sqrt{2 \Omega_J}} (\Omega_J \, \hat{\varphi}_J + {\rm i} \epsilon_J \hat{\pi}_{-J}) \,,  \qquad \qquad \hat{a}_J^\dagger = \frac{1}{\sqrt{2 \Omega_J}}(\epsilon_J \Omega_J \, \hat{\varphi}_{-J} - {\rm i} \hat{\pi}_J) \,,
\end{equation}
with $\Omega_J = \sqrt{|K^{(0)}_J K^{(2)}_J|}$ and $\epsilon_J=(-1)^{\sum_I(j_I-m_I)}$. By construction these operators satisfy
\be
\left[ \hat{a}_J(\chi) , \hat{a}^\dagger_{J'}(\chi)  \right] = \delta_{JJ'}\,,
\ee
and again generate a Fock space (when all $J$ modes are taken into account) by acting on a ground state $|0\rangle$. If we only include modes for which $K_J^{(0)}$ and $K_J^{(2)}$ have opposite sign, the Hamiltonian is  (cf.~appendix \ref{AppA} for the more general case)
\be\label{sqH}
\hat{{H}} = \frac{1}{2} \sum_J \omega_J \left( \hat{a}^\dagger_J \hat{a}^\dagger_{-J} + \hat{a}_J \hat{a}_{-J}\right)
\ee
with $\omega_J = -{\rm sgn}\big( K_J^{(0)}\big) \sqrt{| K^{(0)}_J/K^{(2)}_J|}$. This case of a squeezing Hamiltonian is particularly relevant for applications to cosmology, since squeezing leads to an exponentially growing number of particles under evolution in $\chi$ \cite{toy}, which in turn means the volume for a single mode satisfies the correct effective Friedmann equation for {\em any} state in the theory \cite{Gielen_2020}, in contrast with the focus on very specific states in the algebraic approach. The particle number in modes with harmonic oscillator Hamiltonians is conserved, interpreted as a volume that remains constant in time, so that these modes will eventually be subdominant and are often neglected.

Still working in the Heisenberg picture, a time-dependent number operator is defined as
\begin{equation}\label{Ndep}
	\hat{N}(\chi) = \sum_J \hat{a}_J^\dagger(\chi) \hat{a}_J(\chi)\,,
\end{equation}
and similarly the volume operator as $\hat{V}(\chi) = \sum_J v_J \hat{a}_J^\dagger(\chi) \hat{a}_J (\chi)$. The main difference with \eqref{N} is that \eqref{Ndep} evolves in $\chi$ as dictated by \eqref{Heisen}, i.e., as
\begin{equation}\label{finalresult}
	\hat{N}(\chi) = \hat{U}^\dagger(\chi) \, \hat{N}(0) \, \hat{U} (\chi)\,, \qquad \hat{U}(\chi) = e^{ -{\rm i} \hat{{H}} \chi}\,.
\end{equation}

While this quantisation is straightforward to obtain and interpret and the connection to the classical theory is clear throughout, one might raise the concern that $\chi$ appears only as a ``classical'' parameter with no quantum operators or fluctuations associated to it. This is a common concern with deparametrised approaches, as the expected covariance of the GFT formalism -- the freedom to choose an arbitrary time parameter to express dynamics -- might be broken by making a classical clock choice before quantisation (see, e.g., the general criticism of ``tempus ante quantum'' in \cite{Isham:1992ms,Kuchar:1991qf} and a more specific discussion for GFT in \cite{Marchetti2021}). Our main goal in this paper is to resolve these concerns by embedding the deparametrised approach into a more covariant setting which allows for an arbitrary choice of evolution parameter.

\section{Parametrisation of classical group field theory}\label{ClassicalSec}

In this section we parametrise the classical GFT defined by the action \eqref{DepAction} (for the general idea of parametrisation see, e.g., \cite{Diracbook,Gaugebook}). To start with a simple case, we restrict the formalism to a single field mode $J$ with vanishing magnetic indices, and we postpone the discussion of a quantum theory with multiple field modes to section \ref{MultimodeSec}. The action then reads
\begin{equation}\label{S0GFT}
	S_0[\varphi] =  \frac{1}{2}\sgn(K_J^{(0)})\int {\rm d} \chi  \left[  |K_J^{(0)}| \varphi_J^2 +  |K_J^{(2)}|(\partial_\chi \varphi_J)^2 \right] \,,
\end{equation}
where we chose a mode for which $K_J^{(0)}$ and $K_J^{(2)}$ have opposite sign, meaning the dynamics for this mode are governed by a Hamiltonian of the form (cf.\ \eqref{relham})
\begin{equation}\label{Ham}
	{H}_J (\varphi_J, \pi_J) = \frac{1}{2} \sgn(K_J^{(0)}) \left( \frac{{\pi_J}^2}{|K_J^{(2)}|} - |K_J^{(0)}| \varphi_J^2 \right) \,.
\end{equation}
Because we deal with this (specific) mode only, we drop the label $J$ in the following discussion. Since the global sign of \eqref{Ham} is irrelevant (see appendix \ref{AppA} for details on the eigenvalue problem for the Hamiltonian \eqref{Ham}), from now on we choose $K^{(2)}<0$ and $K^{(0)}>0$ without loss of generality. 

Following the standard parametrisation strategy \cite{Diracbook,Gaugebook}, we now introduce an arbitrary parameter $\tau$ to describe $\varphi(\chi)$ by means of two functions $\varphi(\tau)$ and $\chi(\tau)$, so that the ``group field'' and the matter field are treated parametrically on the same footing. In this manner we obtain a new action 
\begin{equation}\label{S1GFT}
	S [\varphi,\chi]= \frac{1}{2} \int {\rm d}\tau \left[  |K^{(0)}|\varphi^2 \, \partial_\tau \chi + |K^{(2)}| \frac{(\partial_\tau \varphi)^2}{\partial_\tau \chi} \right] \,.
\end{equation}
The Hamiltonian theory derived from the parametrised action \eqref{S1GFT} has an extended phase space spanned by coordinates ($\varphi$, $\chi$) and conjugate momenta ($\pi_\varphi$, $p_\chi$) defined as usual,
\begin{equation}\label{momenta}
	\begin{aligned}
		\pi_\varphi& := \frac{\partial \mathcal{L}}{\partial (\partial_\tau \varphi)} = |K^{(2)}| \frac{\partial_\tau \varphi}{\partial _\tau \chi} \,, \\ 
		p_\chi& := \frac{\partial  \mathcal{L}}{\partial (\partial_\tau \chi)} = \frac{1}{2}|K^{(0)}| \varphi^2- \frac{1}{2}|K^{(2)}| \frac{(\partial_\tau \varphi)^2}{(\partial_\tau \chi)^2} \,.
	\end{aligned}
\end{equation}
It is easy to check that the Hamiltonian associated to the new action \eqref{S1GFT} vanishes. Indeed, the momenta \eqref{momenta} form a constraint
\begin{equation}\label{C}
	C =  p_\chi + H (\varphi, \pi_\varphi)  =0 \,,
\end{equation}
where $H (\varphi, \pi_\varphi)$ is given in \eqref{Ham}. \eqref{C} defines a constraint hypersurface in the extended phase space and generates trajectories on such surface according to 
\begin{equation}\label{classicaldyn}
	\partial_\tau f = \{ f, N C \} \,,
\end{equation}
for any phase space function $f$, where $N$ is a Lagrange multiplier defining the particular parametrisation of these trajectories. From \eqref{classicaldyn} one can find the equations of motion
\begin{equation}\label{ceom}
	\begin{aligned}
		\partial_\tau \varphi &=	\left\{\varphi, NC \right\}=   N \frac{\pi_\varphi}{|K^{(2)}|} \,,\\
		\partial_\tau \chi 	& = \left\{\chi, NC \right\} = N \,,\\
	\end{aligned}
\end{equation}
while $	\partial_\tau \pi_\varphi =\partial_\tau p_\chi =0$. Of course, combining the equations \eqref{ceom} one finds the definition of $\pi_\varphi$ in \eqref{momenta}. The second equation in \eqref{ceom} shows that $N$ gives the rate of change of $\chi$ with respect to the label $\tau$, and it is sometimes called lapse function for this reason (borrowing the nomenclature from general relativity). Using \eqref{momenta}, \eqref{C} and \eqref{ceom} one can formulate the same dynamics starting from the action
\begin{equation}\label{S1}
	S [\varphi,\chi]= \int \dd \tau \left( \pi_\varphi \partial_\tau \varphi +p_\chi \partial_\tau \chi - N C \right)\,,
\end{equation}
which explicitly shows that $NC$ plays the role of the Hamiltonian (sometimes called super-Hamiltonian), and indeed yields the same equations of motion \eqref{ceom}. Note that such a parametrised theory describes the same physics of the initial action $S_0[\varphi]$ (cf.\ \eqref{S0GFT}); while \eqref{S1} contains one extra canonical pair ($\chi$, $p_\chi$), it also implies the constraint \eqref{C} (obtained by varying with respect to $N$). Since \eqref{C} is a first-class constraint, it eliminates two degrees of freedom so that the two actions describe the same number of independent degrees of freedom. We have now introduced a form of ``general covariance'', as \eqref{S1} is invariant under $\tau$-reparametrisation. This symmetry is reflected in the fact that $N$ can be an arbitrary function, playing the role of a gauge field.

The action \eqref{S1} is comprised of two parts: one related to geometry described by the group field $\varphi$, and one for the matter scalar field $\chi$. Since we will want to use the scalar field $\chi$ as internal dynamical clock to describe the GFT system (which corresponds to the choice of the lapse $N=1$), we briefly review here the notions of classical relational dynamics and Dirac observables. Following \cite{DittrichDO,GieselDO,Tambornino}, we begin by noticing that \eqref{classicaldyn} defines a flow $\alpha_C^\tau$ with parameter $\tau$ that transforms a phase space function $f$ as
\begin{equation}\label{flow}
	f \mapsto \alpha_C^\tau (f) := \sum_{n=0}^\infty \frac{\tau^n}{n!} \{f, C\}_n \,,
\end{equation}
where $\{f, C\}_n$ denotes the iterated Poisson bracket defined as $\{f, C\}_{n+1} := \{\{f, C\}, C\}_n$ with $\{f, C\}_0=f$. Given that the action \eqref{S1} is invariant under $\tau$-reparametrisation, the evolution with respect to the flow parameter $\tau$ is not physical as it is a gauge transformation on the constraint hypersurface defined by \eqref{C}. Physical observables (known as Dirac observables) are defined as functions of canonical variables  $F(\varphi,\pi_\varphi,\chi,p_\chi)$ that are invariant under $\tau$ evolution. Then, they satisfy
\begin{equation}\label{classicalFC}
	\{F, C\} \approx 0 \,,
\end{equation}
where $\approx$ represents a ``weak equality'', meaning the equality holds on the constraint hypersurface. In other words, functions $F$ satisfying \eqref{classicalFC} are constant along the trajectories (within the constraint hypersurface) generated by the constraint \eqref{C}. 

One can now use the strategy of ``evolving constants of motion'' \cite{Rovelli_Amodel, Rovelli_Anhypothesis,RovelliOBS,RovelliPO} to define \textit{relational Dirac observables}, which evolve with respect to another chosen observable along the flow generated by $C$. These are also known as ``complete'' observables $F_{f,\chi}(\chi_0)$, and correspond to the value a partial observable $f$ takes on $C$ when another partial observable $\chi$ takes the value $\chi_0$. The second partial observable $\chi$ is thought as a dynamical ``clock'' degree of freedom, chosen to parametrise the flow in place of the unphysical parameter $\tau$. In short, one can construct a complete observable satisfying \eqref{classicalFC} as \cite{DittrichDO,GieselDO,Tambornino,Trinity,RelativisticTrinity}
\begin{equation}\label{genericDO}
	F_{f,\chi}(\chi_0)   \approx \sum_{n=0}^\infty \frac{(\chi_0-\chi)^n}{n!} \left\{ f , \frac{C}{\{\chi , C\} }\right\}_n \,,
\end{equation}
which is well-defined if $\{\chi , C\} \neq 0$ (i.e., if $\chi$ is a good clock to parametrise the flow). While \eqref{genericDO} holds for (finite-dimensional) systems with a generic Hamiltonian constraint \cite{DittrichDO,GieselDO,Tambornino}, our scenario belongs to the specific class of systems thoroughly analysed in \cite{Trinity,RelativisticTrinity}; a simplification arises because of the partition of the classical constraint \eqref{C} into a $\chi$ component and a $\varphi$ component
\begin{equation}\label{CHHclass}
	C =  H_\chi +H_\varphi  =  0\,,
\end{equation}
where the so-called clock Hamiltonian is $H_\chi = p_\chi$ and  the GFT Hamiltonian $H_\varphi$ is given in \eqref{Ham}. Thanks to the crucial fact that the clock Hamiltonian is canonically conjugate to $\chi$, 
\begin{equation}\label{chipichiclass}
	\{\chi, H_\chi\} = \{\chi, p_\chi\} = 1\,,
\end{equation} 
one can show that a relational Dirac observable associated to a function $f_\varphi$ of the GFT phase space (i.e., a function of $\varphi$ and $\pi_\varphi$) takes the simple form \cite{DittrichDO,GieselDO,Tambornino,Trinity,RelativisticTrinity}
\begin{equation}\label{classicalDO}
	F_{f_\varphi,\chi} (\chi_0) \approx \sum_{n=0}^\infty \frac{(\chi_0-\chi)^n}{n!} \{f_\varphi, H_\varphi\}_n \,.
\end{equation}
The partition into a matter clock sector and a geometry sector will be exploited in the quantum theory to express the dynamics in the general framework of \cite{Trinity,RelativisticTrinity}; this will allow to relate the parametrised theory introduced here to existing approaches to canonical quantisation of GFT.

\section{Quantum theory for single mode}\label{QuantumSec}

\subsection{Kinematics}

We now quantise the theory described by $S [\varphi,\chi]$ in \eqref{S1}, promoting the canonical coordinates and momenta to operators with the commutation relations
\begin{equation}\label{commutators}
	\left[\hat{\varphi}, \hat{\pi}_\varphi\right] = {\rm i}  \,, \qquad \qquad
	\left[\hat{\chi}, \hat{p}_\chi\right] = {\rm i}  \,,
\end{equation}
all the others being zero. Kinematically, the Hilbert space of such a quantum theory is the tensor product of two Hilbert spaces: one for the matter sector associated with the scalar field $\chi$ and one for the geometry sector associated with the group field $\varphi$, namely
\begin{equation}\label{Hilbert}
	\mathcal{H}_{\text{kin}} = \mathcal{H}_\chi \otimes \mathcal{H}_\varphi \,.
\end{equation}
Both $\mathcal{H}_\chi $ and $\mathcal{H}_\varphi$ are spaces of square-integrable functions over the real line, so that $\mathcal{H}_{\text{kin}}=L^2(\mathbb{R}^2)$.

\textbf{Geometry sector.} The Hamiltonian for the (single-mode) GFT system living on the geometry sector is given by promoting \eqref{Ham} to an operator on $\mathcal{H}_\varphi$ as
\begin{equation}\label{GFTH}
	\hat{H}_\varphi = \frac{1}{2} \left( \frac{\hat{\pi}_\varphi^2}{|K^{(2)}|} - |K^{(0)}| \hat{\varphi}^2 \right)\,,
\end{equation}
which resembles the Hamiltonian of a quantum particle with an inverted harmonic potential. The Schr\"odinger problem for the Hamiltonian \eqref{GFTH} can be solved explicitly \cite{Damped1,Damped2} (we refer to appendix \ref{AppA} for the details) to find a doubly degenerate continuous energy spectrum
\begin{equation}\label{energydouble}
	\hat{H}_\varphi |\psi_\pm^E \rangle= E |\psi_\pm^E \rangle\,,
\end{equation}
where the energy eigenstates can be expressed in terms of special functions known as parabolic cylinder functions \cite{abramowitzstegun,GradRy}. Two essential properties that we will need when discussing relational quantum dynamics in section \ref{ReldynSec} are the generalised orthonormality condition
\begin{equation}\label{orthoGFT}
	\langle \psi_m^E |\psi_n^{E'}\rangle = \delta_{mn} \delta (E-E')\,,
\end{equation}
where $m$ and $n$ can either be $+$ or $-$, and the spectral resolution of the Hamiltonian \eqref{GFTH}
\begin{equation}\label{spectrumGFT}
	\hat{H}_\varphi =\int_\pm \dd E \; E \; |\psi_\pm^E \rangle \langle \psi_\pm^E|\,,
\end{equation}
where the notation $\int_\pm   := \sum_\pm \int $ takes into account the double degeneracy of the spectrum (see appendix \ref{AppA}). Note that while a Schr\"odinger equation for GFT with such a Hamiltonian was introduced in \cite{relham_Wilson_Ewing_2019,relhamadd}, there was no discussion of exact solutions (even for a single field mode). 

Similarly to the deparametrised approach described in section \ref{GFTSec}, one can now change basis and introduce ladder operators $\hat{\mathfrak{a}}$ and $\hat{\mathfrak{a}}^\dagger$. These are to be considered \textit{kinematical} operators at this stage: the dynamics of our (parametrised) theory are only defined when the quantum version of the constraint \eqref{CHHclass} is used, as we will do in the next section. In other words, one can build a \textit{kinematical} Fock space starting from the Fock vacuum $|0\rangle$ defined by $\hat{\mathfrak{a}} |0\rangle =0 $, and building $n$-particle states in the usual way; for example, the one-particle state reads $|1\rangle =\hat{\mathfrak{a}}^\dagger |0\rangle$. Moreover, one can discuss kinematical operators built from $\hat{\mathfrak{a}}$ and $\hat{\mathfrak{a}}^\dagger$, such as the number operator
\begin{equation}\label{Npar}
	\hat{\mathfrak{N}}= \hat{\mathfrak{a}}^\dagger \hat{\mathfrak{a}}  \,,
\end{equation}
and the volume operator $\hat{\mathfrak{V}} = v \hat{\mathfrak{N}}$. We use a new notation for these kinematical operators to emphasise that these are \textit{a priori} different operators than the ones introduced in the deparametrised approach.

\textbf{Matter sector.} As already discussed at the classical level, we will want to use the matter field as relational clock. The quantum theory living on $\mathcal{H}_\chi$ is isomorphic to the Hilbert space of a particle on a line, so we have the following properties for the $\hat{\chi}$ operator
\begin{equation}\label{chiproperties}
	\begin{aligned}
		\hat{\chi} |\chi\rangle & = \chi |\chi\rangle \,, \\
		\langle \chi |\chi'\rangle &= \delta(\chi-\chi') \,,
	\end{aligned}
\end{equation}
and similarly for its conjugate momentum
\begin{equation}\label{pchiproperties}
	\begin{aligned}
		\hat{p}_\chi |p_\chi \rangle & = p_\chi |p_\chi\rangle \,, \\
		\langle p_\chi |p'_\chi \rangle &= \delta(p_\chi-p'_\chi) \,,
	\end{aligned}
\end{equation}
where we choose conventions with $\langle p_\chi | \chi\rangle = \frac{1}{\sqrt{2\pi}} e^{-{\rm i} p_\chi \chi}$. It follows that one can write the identity on $\mathcal{H}_\chi$ as
\begin{equation}\label{resolutionI}
	\mathbb{I}_\chi = \int \dd \chi |\chi\rangle \langle\chi | = \int \dd p_\chi |p_\chi \rangle \langle p_\chi |\,,
\end{equation}
and the spectral decomposition of $\hat{\chi}$ and $\hat{p}_\chi$ as
\begin{align}
	\hat{\chi} &= \int \dd \chi \; \chi\; | \chi \rangle \langle \chi \label{timeoperator} |\,,\\
	\hat{H}_\chi &= \hat{p}_\chi = \int \dd p_\chi \; p_\chi\; | p_\chi \rangle \langle p_\chi  |\,.
\end{align}

Notice that interpreting $\hat{p}_\chi$ as the (quantum) clock Hamiltonian (cf.\ \eqref{CHHclass} and \eqref{chipichiclass}) implies that the operator $\hat{\chi}$ in \eqref{timeoperator} will be interpreted as time operator.\footnote{More formally, one can define the time operator $\hat{\chi}$ as the first moment operator (cf.\ \eqref{timeoperator}) of the ``time observable'' $E_\chi:=|\chi\rangle\langle\chi|$, in turn defined via the most general notion of quantum observable as a positive operator-valued measure (POVM) \cite{Trinity,RelativisticTrinity}.} The clock states correspond to eigenstates $|\chi\rangle$ of the time operator $\hat{\chi}$. Such states ``evolve'' under the action of the group generated by $\hat{p}_\chi$ as
\begin{equation}\label{evolvingtime}
	\hat{U}_\chi (\alpha)|\chi'\rangle = 	|\chi'+\alpha\rangle\,,
\end{equation}
where
\begin{equation}\label{lastchiprop}
	\hat{U}_\chi (\alpha): = e^{-{\rm i} \hat{p}_\chi \alpha}\,, \qquad \alpha \in \mathbb{R}\,.
\end{equation}

\subsection{Relational dynamics}\label{ReldynSec}

One now makes use of the quantum version of the classical constraint \eqref{C}, $\hat{C}$, to identify physical states among the kinematical ones. This is the first step of the Dirac quantisation procedure, which allows to discuss the notion of relational Dirac observables in a precise sense (via the quantum analogue of \eqref{classicalFC} and \eqref{classicalDO}). At the same time, the tensor product structure of \eqref{Hilbert} suggests that the Page--Wootters formalism could also be used to implement the notion of relational dynamics. For this reason, we explicitly write the quantum constraint as
\begin{equation}\label{Cquantum}
	\hat{C} =  \hat{p}_\chi +\hat{H}_\varphi  = \hat{p}_\chi\otimes \mathbb{I}_\varphi + \mathbb{I}_\chi \otimes \hat{H}_\varphi  \,,
\end{equation}
where $\mathbb{I}_\varphi$ and $\mathbb{I}_\chi$ are the identity operators on $ \mathcal{H}_\varphi$ and $\mathcal{H}_\chi $ respectively. It was shown in \cite{Trinity,RelativisticTrinity} that with a constraint of the form \eqref{Cquantum}, the Dirac algorithm for a constraint quantisation (for instance implemented using group averaging techniques) and the Page--Wootters formalism yield equivalent relational dynamics, as we will exemplify with our GFT model. 

From now on, both operators and states will occasionally have a subscript $\chi$ or $\varphi$, to clarify (when necessary) in which sector of \eqref{Hilbert} they act or live.

\textbf{Dirac quantisation.} A generic state in the kinematical Hilbert space \eqref{Hilbert}, $|\Psi_{\text{kin}} \rangle \in \mathcal{H}_{\text{kin}}$, can be written as
\begin{equation}\label{kinstate}
	|\Psi_{\text{kin}}\rangle =  \int_\pm \dd E \int \dd p_\chi \,  \Psi_\pm (p_\chi, E) \, |p_\chi \rangle_\chi \otimes | \psi^E_\pm\rangle_\varphi\,.
\end{equation}
Following the Dirac programme for quantising constrained systems, one defines physical states by demanding that they are annihilated by the constraint \eqref{Cquantum},
\begin{equation}\label{Cpsizero}
	\hat{C} |\Psi_{\text{phys}}\rangle =0 \,.
\end{equation}
As is well known \cite{ThiemannBook}, such physical states are not normalisable in $\mathcal{H}_{\text{kin}}$; one needs to introduce a new inner product since $\langle \Psi_{\text{phys}}|	\Psi_{\text{phys}}\rangle_{\text{kin}}$ diverges, where $\langle \cdot |\cdot \rangle_{\text{kin}}$ is the inner product on \eqref{Hilbert}. One way of doing this is by ``projecting'' a kinematical state onto a physical one by means of group averaging \cite{ThiemannBook,Marolf1,Marolf2},
\begin{equation}\label{deltaC}
	\delta (\hat{C}) = \frac{1}{2\pi } \int \dd \alpha \, e^{{\rm i} \alpha \hat{C}}\,,
\end{equation}
as $|\Psi_{\text{phys}}\rangle = \delta (\hat{C}) |\Psi_{\text{kin}}\rangle$, and then defining a physical inner product as
\begin{equation}\label{PIP}
	\langle \Psi_{\text{phys}}|	\Psi_{\text{phys}}\rangle_{\text{phys}} := \langle \Psi_{\text{kin}}| \delta (\hat{C})  |	\Psi_{\text{kin}}\rangle_{\text{kin}} \,.
\end{equation}
Starting from \eqref{kinstate} and using properties \eqref{energydouble}, \eqref{orthoGFT} and \eqref{pchiproperties}, one explicitly finds
\begin{equation}\label{phys1m}
	|\Psi_{\text{phys}}\rangle = \int_\pm \dd E \, \Psi_\pm (-E,E) \;|-E\rangle_\chi \otimes |\psi_\pm^E\rangle_\varphi \,,
\end{equation}
and
\begin{equation}\label{Physnorm}
	\langle \Psi_{\text{phys}}|	\Psi_{\text{phys}}\rangle_{\text{phys}} =  \int_\pm \dd E \, |\Psi_\pm (-E,E)|^2 \,.
\end{equation}
This norm defines $\mathcal{H}_\text{phys}$ as the space of solutions to \eqref{Cpsizero}. Physical states do not change under the flow of the total Hamiltonian $\hat{C}$,
\begin{equation}\label{timeless}
	\hat{U}_{\chi\varphi}(\alpha) |\Psi_\text{phys}\rangle = |\Psi_\text{phys}\rangle \,,
\end{equation}
with
\begin{equation}\label{Uchiphi}
	\hat{U}_{\chi\varphi}(\alpha) := e^{-{\rm i}\alpha \hat{C}} = e^{-{\rm i} \alpha \hat{p}_\chi} \otimes e^{-{\rm i} \alpha \hat{H}_\varphi} \,, \qquad \alpha \in \mathbb{R}\,,
\end{equation}
and are sometimes called ``timeless'' or ``frozen''. More recently \cite{Trinity,RelativisticTrinity}, they have been denoted ``clock-neutral'' as they describe physics before choosing a temporal reference system. \eqref{timeless} is what gives rise to the {problem of time} \cite{Isham:1992ms,Kuchar:1991qf}, which can be tackled by defining the quantum counterpart of Dirac relational observables (cf.\ \eqref{classicalFC} and \eqref{classicalDO}). Indeed, one can now choose the temporal reference system (namely, the clock) associated with the Hilbert space $\mathcal{H}_\chi$ with properties \eqref{chiproperties}--\eqref{evolvingtime}, and find the quantised version of \eqref{classicalDO} as \cite{Trinity,RelativisticTrinity}
\begin{equation}\label{Fquantum}
	\hat{F}_{f_\varphi,\chi}(\chi_0) = \frac{1}{2\pi} \int \dd \chi\, | \chi \rangle\langle \chi | \otimes \sum_{n=0}^\infty \frac{{\rm i}^n}{n!} (\chi-\chi_0)^n \left[\hat{f}_\varphi , \hat{H}_\varphi \right]_n \,,
\end{equation} 
where the commutator $[\hat{f}_\varphi , \hat{H}_\varphi ]_n : = [[\hat{f}_\varphi , \hat{H}_\varphi ]_{n-1} , \hat{H}_\varphi]$ and $[\hat{f}_\varphi , \hat{H}_\varphi ]_0 := \hat{f}_\varphi$. Thanks to the Baker-Campbell-Hausdorff formula, \eqref{Fquantum} can be equivalently recast as
\begin{equation}\label{Fquantumuseful}
	\begin{aligned}
		\hat{F}_{f_\varphi,\chi}(\chi_0) &= \frac{1}{2\pi} \int \dd \chi \, |\chi\rangle \langle \chi| \otimes \hat{U}_\varphi (\chi-\chi_0)  \hat{f}_\varphi \hat{U}_\varphi^\dagger(\chi-\chi_0)\\
		& = \frac{1}{2\pi} \int \dd \alpha  \, \hat{U}_{\chi\varphi} (\alpha) \left( |\chi_0\rangle\langle \chi_0 | \otimes \hat{f}_\varphi \right) \hat{U}_{\chi\varphi}^\dagger(\alpha) \,,
	\end{aligned}
\end{equation}
where $\hat{U}_\varphi (\alpha) := e^{-{\rm i} \hat{H}_\varphi \alpha}$, $\hat{U}_{\chi\varphi}(\alpha)$ is given in \eqref{Uchiphi}, and in the last line we changed integration variable, $\chi\rightarrow \alpha+\chi_0$. Crucially, quantum relational observables defined using the prescription \eqref{Fquantum} commute with the constraint operator $\hat{C}$,
\begin{equation}\label{CFquantum}
	\left[ \hat{F}_{f_\varphi,\chi}(\chi_0) , \hat{C} \right] =0\,,
\end{equation}
and are thus called (quantum) Dirac observables.\footnote{Since our clock Hamiltonian is simply $\hat{H}_\chi = \hat{p}_\chi$, the commutator in \eqref{CFquantum} vanishes strongly (i.e., algebraically). In \cite{Trinity,RelativisticTrinity} it is shown that for more complicated clock Hamiltonians one can still prove that a Dirac observable associated to a {physical} phase space function weakly commutes with the constraint $\hat{C}$, namely when applied to physical states.} 

The prototype Dirac observable for our GFT model is the number operator $\hat{f}_\varphi = \hat{\mathfrak{N}}$ (cf.\ \eqref{Npar}); thus, we will be specifically interested in the observable
\begin{equation}\label{DiracN}
	\hat{{N}}_D(\chi_0) := \frac{1}{2\pi} \int \dd \alpha \, \hat{U}_{\chi\varphi} (\alpha) \left( |\chi_0\rangle \langle \chi_0 | \otimes\hat{\mathfrak{N}} \right) \hat{U}_{\chi\varphi}^\dagger(\alpha) \,,
\end{equation}
where the subscript $D$ refers to the fact that this is a Dirac observable on the total Hilbert space. As opposed to the kinematical counterpart $\hat{\mathfrak{N}}$, the operator $\hat{{N}}_D(\chi_0)$ evolves as the parameter $\chi_0$ runs (taking the values the time operator $\hat{\chi}$ can take), meaning that \eqref{DiracN} truly defines relational quantum dynamics on $\mathcal{H}_{\text{phys}}$ for the GFT number operator. In particular, even if physical states do not transform under the action of the Hamiltonian constraint $\hat{C}$, we will evaluate the relational Dirac observable \eqref{DiracN} using the physical inner product \eqref{PIP} so to obtain an expectation value for the number operator which indeed changes with respect to the matter scalar field.

\textbf{Page--Wootters formalism.} The framework introduced by Page and Wootters \cite{PW,Wootters} provides another way to define relational dynamics for systems subject to a quantum constraint of the form \eqref{Cquantum}. Specifically, starting again with a kinematical Hilbert space \eqref{Hilbert} that is split into clock and system (our matter scalar field $\chi$ and single-mode GFT model, respectively), one selects physical states using the constraint equation \eqref{Cpsizero}. The apparent difference with the previous section arises when choosing an inner product on $\mathcal{H}_{\text{phys}}$ to completely specify the space of solutions of \eqref{Cpsizero}. 

The conceptual idea behind the Page--Wootters formalism is to interpret quantum theory with conditional probabilities. More precisely, one defines the state of a system at a given instant of (relational) time as a solution to the constraint equation \textit{conditioned} on a subsystem of the theory to be in a state corresponding to that time. Following this idea, we define the state of our single-mode GFT system at a given time, say $\chi_0$, as a solution to the constraint \eqref{Cpsizero} conditioned on the clock being in the state $|\chi_0\rangle$,
\begin{equation}\label{conditioned}
	|\psi(\chi_0)\rangle_\varphi  : = \Big(\langle \chi_0| \otimes \mathbb{I}_\varphi \Big) |\Psi_{\text{phys}}\rangle\,.
\end{equation}
It follows that a physical state can be expressed as a ``history state'', namely
\begin{equation}\label{history}
	|\Psi_{\text{phys}}\rangle = \int \dd \chi_0 \, |\chi_0\rangle_\chi \otimes |\psi(\chi_0)\rangle_\varphi \,,
\end{equation}
since $\mathbb{I}_\chi \otimes \mathbb{I}_\varphi = \int \dd \chi_0 |\chi_0\rangle\langle \chi_0| \otimes \mathbb{I}_\varphi$. Indeed, the state \eqref{history} encodes information about the whole timeline. The usual formulation of quantum mechanics for the conditioned states can be recovered in terms of a Schr\"odinger equation in the clock time $\chi_0$ \cite{PW,Wootters},
\begin{equation}\label{Schro}
	{\rm i}  \frac{\dd}{\dd \chi_0}|\psi(\chi_0)\rangle _\varphi = \hat{H}_\varphi |\psi(\chi_0)\rangle_\varphi\,,
\end{equation}
which is easily found from the constraint equation as $\Big({_\chi}\langle \chi_0| \otimes \mathbb{I}_\varphi \Big) \hat{C} |\Psi_{\text{phys}}\rangle =0 $. Moreover, one introduces what is known as the Page--Wootters inner product \cite{SmithPW,Trinity,RelativisticTrinity}
\begin{equation}\label{PWIP}
	\langle \Psi_\text{phys}|\Psi_\text{phys}\rangle _{\text{PW}}  : = \langle \Psi_\text{phys}|\Big(|\chi_0 \rangle \langle \chi_0 | \otimes \mathbb{I}_\varphi \Big)  |\Psi_\text{phys}\rangle_\text{kin}\,,
\end{equation}
which is consistent with the usual inner product on $\mathcal{H}_\varphi$ at all times since
\begin{equation}
	\begin{aligned}
		\langle \Psi_\text{phys}|\Psi_\text{phys}\rangle _{\text{PW}} &= \left(\int \dd \chi' \, {}_\chi\langle \chi' | \otimes {}_\varphi \langle \psi(\chi') | \right)  \Big(|\chi_0 \rangle_\chi {}_\chi \langle \chi_0 | \otimes \mathbb{I}_\varphi \Big) \left(\int \dd \chi \, | \chi \rangle_\chi \otimes | \psi(\chi) \rangle_\varphi \right)\\
		& = {}_\varphi\langle \psi (\chi_0)|\psi(\chi_0) \rangle_\varphi \,,
	\end{aligned}
\end{equation}
where ${}_\varphi\langle \psi (\chi_0)|\psi(\chi_0) \rangle_\varphi = {}_\varphi\langle \psi (0)|\psi(0) \rangle_\varphi$ is independent of $\chi_0$ thanks to \eqref{Schro} and the fact that $\hat{H}_\varphi$ is self-adjoint (cf.\ \eqref{GFTH}).   

In what follows we will provide an application of the equivalence shown in \cite{Trinity,RelativisticTrinity} between the relational dynamics defined using a Dirac quantisation and using the Page--Wootters formalism, mainly focussing on the number operator of our GFT model. To begin with, it is easy to check explicitly that
\begin{equation}\label{PWforus}
	\langle \Psi_\text{phys} | \Psi_\text{phys} \rangle_{\text{PW}} = \int_\pm \dd E \, |\Psi_\pm (-E,E)|^2
\end{equation} 
is the same as \eqref{Physnorm}. Note that in the following calculations we use the generic expressions for the physical inner product \eqref{PIP} and the Page--Wootters inner product \eqref{PWIP} instead of their explicit form \eqref{PWforus} (or \eqref{Physnorm}). We can now turn to the calculation of the expectation value of the Dirac observable associated to the number operator \eqref{DiracN}, using the physical inner product.  First, we find
\begin{equation}\label{Nonpsi}
	\begin{aligned}
		\hat{{N}}_D(\chi_0) |\Psi_\text{phys} \rangle  & = \frac{1}{2\pi} \int \dd \alpha \, \hat{U}_{\chi\varphi} (\alpha) \left( |\chi_0\rangle  \langle \chi_0 | \otimes\hat{\mathfrak{N}} \right) |\Psi_\text{phys} \rangle\\
		& = \frac{1}{2\pi} \int \dd \alpha \, \hat{U}_\chi (\alpha) |\chi_0\rangle \langle \chi_0| \otimes \hat{U}_\varphi(\alpha) \hat{\mathfrak{N}} |\Psi_\text{phys} \rangle\\ 
		& = 	\delta(\hat{C}) \left( |\chi_0 \rangle \langle \chi_0| \otimes \hat{\mathfrak{N}}  \right) |\Psi_\text{phys} \rangle \,,
	\end{aligned}
\end{equation}
where we used $\hat{U}_{\chi\varphi}^\dagger(\chi) |\Psi_\text{phys} \rangle =|\Psi_\text{phys} \rangle $ (cf.\ \eqref{timeless}) and the definitions \eqref{Uchiphi} and \eqref{deltaC}. Then, we calculate the expectation value in the physical inner product \eqref{PIP} as
\begin{equation}\label{equivalenceN}
	\begin{aligned}
		N_D(\chi_0) & := 	\langle \Psi_\text{phys} | \hat{{N}}_D (\chi_0) |\Psi_\text{phys} \rangle_\text{phys} \\&= \langle \Psi_\text{phys} | \delta(\hat{C}) \left( |\chi_0 \rangle \langle \chi_0| \otimes \hat{\mathfrak{N}}  \right) |\Psi_\text{phys} \rangle_\text{phys}\\
		& = \langle \Psi_\text{kin} | \delta(\hat{C}) \left( |\chi_0 \rangle \langle \chi_0| \otimes \hat{\mathfrak{N}}  \right) \delta(\hat{C}) |\Psi_\text{kin} \rangle_\text{kin}\\
		& =\langle \Psi_\text{phys} | \left( |\chi_0 \rangle \langle \chi_0| \otimes \hat{\mathfrak{N}}  \right)  |\Psi_\text{phys} \rangle_\text{kin}\\
		& = \langle \Psi_\text{phys} | \hat{\mathfrak{N}}    |\Psi_\text{phys} \rangle_\text{PW}\,.
	\end{aligned}
\end{equation}
Of course, using the definition of conditioned states \eqref{conditioned}, it is easy to show that \eqref{equivalenceN} gives back the result of the deparametrised approach of section \ref{GFTSec}:
\begin{equation}\label{result}
	\begin{aligned}
		N_D(\chi_0) &= \langle \Psi_\text{phys} | \left( |\chi_0 \rangle \langle \chi_0| \otimes \hat{\mathfrak{N}}  \right)  |\Psi_\text{phys} \rangle_\text{kin}\\
		&=  {}_\varphi \langle \psi(\chi_0)| \hat{\mathfrak{N}}  | \psi(\chi_0)\rangle _\varphi\\
		&=  {}_\varphi \langle \psi |\, \hat{U}^\dagger_\varphi (\chi_0)\,  \hat{\mathfrak{N}} \, \hat{U}_\varphi (\chi_0) \, | \psi \rangle _\varphi\,,
	\end{aligned}
\end{equation}
where $|\psi\rangle_\varphi = |\psi(0)\rangle_\varphi$ and we switched from the Schr\"odinger picture on the second line to the Heisenberg picture on the third line. Indeed, \eqref{result} shows that the expectation value of the relational Dirac observable $\hat{N}_D(\chi_0)$, computed with the physical inner product on $\mathcal{H}_\text{phys}$, is equivalent to the expectation value of \eqref{finalresult} (here for a single-mode) in the deparametrised quantum theory. Of course, the same holds true for the volume operator (and any other one built from ladder operators), so that all the GFT results and applications to cosmology are recovered in our parametrised theory. 

The equivalence established in \eqref{result} strengthens the deparametrised approach described in section \ref{GFTSec}, since the relational Dirac observable $\hat{N}_D(\chi_0)$ in \eqref{DiracN} is defined relationally with respect to the eigenvalues of a time operator $\hat{\chi}$, in contrast with the deparametrised theory where the clock label is chosen before quantisation. In this sense, the construction of section \ref{QuantumSec} belongs to the ``tempus post quantum'' type of relational dynamics discussed in \cite{Marchetti2021} (see also \cite{Isham:1992ms,Kuchar:1991qf}).

Moreover, \eqref{equivalenceN} shows that $N_D(\chi_0)$ is also equal to the Page--Wootters expectation values of the corresponding kinematical operator $\hat{\mathfrak{N}}$, which allows to reinterpret GFT observables as conditional on the clock. In particular, this suggests a canonical picture characterised by a splitting between quantum geometry and a constant-time ``slice'' on the history state \eqref{history}, where the internal time takes a fixed value $\chi_0$. This construction is similar in spirit to the $3+1$ splitting of canonical GR, where one describes evolution as a sequence of constant time hypersurfaces. While the physical state \eqref{history} of Page and Wootters is a superposition of all clock states and all GFT states, it allows to answer the relevant questions of what happens at any specific value of relational time. In other words, this approach justifies the interpretation of the GFT quanta as being associated to the same clock reading since the conditioned state \eqref{conditioned}, by definition, involves a projection onto the $\chi_0$ slice. 

\section{Extension to multiple field modes}\label{MultimodeSec}

In this section we show that our construction can be extended to any number of Peter--Weyl modes. While in principle a GFT contains an infinite number of modes $J$ (see \eqref{peterw}), we will consider a truncation to a finite (but arbitrary) number of modes. Such a truncation could be elegantly implemented using the quantum group $SU_q(2)$, where the deformation parameter $q$ is related to a nonvanishing cosmological constant (see, e.g., \cite{Major,SmolinLambda,Freidel98,CarloEugenio}), which leads to a maximum irreducible representation and hence a maximum value for the multi-index $J$. Similar ideas on $q$-deformation were recently applied in the context of three-dimensional group field theories in \cite{GirelliGFTqg}, and provide a way to implement a cut-off value $J_\text{max}$ in the Peter--Weyl decomposition \eqref{peterw}. We assume such a cut-off in the following sections.

\subsection{Single reparametrisation symmetry}\label{MultimodeSecA}

Without making any assumptions on magnetic indices and on the relative sign of $K^{(0)}_J$ and $K^{(2)}_J$ for the various modes, we start the analysis with the generic action given in \eqref{DepAction}. As a first generalisation of the procedure described in section \ref{ClassicalSec}, we proceed here by introducing a parameter $\tau$ for the theory with multiple field modes. Then, all the $\varphi_J(\tau)$'s as well as $\chi(\tau)$ depend on $\tau$, and by means of the chain rule one obtains the action
\begin{equation}\label{supar}
	S[\varphi, \chi] = \frac{1}{2} \int \dd \tau \, \sum_J 
	\left( K^{(0)}_J \varphi_{-J}\varphi_J \, \partial_\tau \chi  -  K_J^{(2)}\frac{(\partial_\tau\varphi_{-J})(\partial_\tau\varphi_J)}{\partial_\tau \chi } \right)\,.
\end{equation}
All the key points discussed for the single-mode scenario also apply here. In particular, while we have enlarged the phase space, the theory is now subject to a Hamiltonian constraint (cf.\ \eqref{C})
\begin{equation}\label{CA}
	C = {H}_\varphi^\text{tot} + p_\chi =0 \,,
\end{equation}
where, rewriting \eqref{relham} here for convenience,
\begin{equation}\label{Htot}
	{H}_\varphi^\text{tot} = -\frac{1}{2} \sum_{J}
	\left[\frac{\pi_{J}(\chi)\pi_{-J}(\chi)}{K^{(2)}_{J}}+K^{(0)}_{J}\varphi_{J}(\chi)\varphi_{-J}(\chi)\right] \,,
\end{equation}
with momenta given by
\begin{align}
	\pi_J &= -K^{(2)}_J \frac{\partial_\tau \varphi_{-J}}{\partial_\tau \chi}  \,, \\
	p_\chi & = \frac{1}{2} \sum_J
	\left( K^{(0)}_J \varphi_{-J}\varphi_J +K^{(2)}_J \frac{(\partial_\tau \varphi_{-J}) (\partial_\tau \varphi_J)}{(\partial_\tau \chi)^2}  \right)\,.
\end{align}
The equations of motion for every group field mode, $\partial_\tau \varphi_J = - N {\pi_{-J}}/{K^{(2)}_J}$, and for the matter field, $\partial_\tau \chi=N $, can be obtained as in \eqref{ceom}. They also follow from the action (in Hamiltonian form)
\begin{equation}\label{Saa}
	S[\varphi, \chi] =\int \dd \tau \, \left( \sum_J
	\pi_J \partial_\tau \varphi_{J} +p_\chi \partial_\tau \chi - N C \right) \,,
\end{equation}
which generalises \eqref{S1} for multiple modes.

Finally, in complete analogy with section \ref{ClassicalSec}, one can adopt the strategy of relational Dirac observables and define quantities satisfying $\{F, C\}\approx 0$ as
\begin{equation}\label{multiRDO}
	F_{f_\varphi,\chi} (\chi_0) \approx \sum_{n=0}^\infty \frac{(\chi_0 -\chi)^n}{n!} \{f_\varphi,{H}_\varphi^\text{tot}\}_n \,,
\end{equation}
where ${H}_\varphi^\text{tot}$ is given in \eqref{Htot} and $f_\varphi$ is now a function of multiple GFT modes. The complete observable $F_{f_\varphi,\chi} (\chi_0)$ associates values of $f_\varphi$ to the specific relational time $\chi=\chi_0$ (see section \ref{ClassicalSec}).

The quantum theory corresponding to the above construction is obtained in the usual manner by means of the commutators
\begin{equation}\label{multiJcomms}
	\left[\hat{\varphi}_J, \hat{\pi}_{J'}\right] = {\rm i} \delta_{JJ'  } \,,\qquad \qquad 		\left[\hat{\chi}, \hat{p}_\chi\right] = {\rm i}\,,
\end{equation}
where the kinematical Hilbert space can be written as
\begin{equation}\label{bigHilbert}
	\mathcal{H}_{\text{kin}}= \mathcal{H}_\chi \otimes \mathcal{H}^{\text{tot}}_\varphi \,, \qquad \qquad   \mathcal{H}^{\text{tot}}_\varphi = \bigotimes_J 
	\mathcal{H}_{\varphi_J}    \,,
\end{equation}
namely as a tensor product of a matter clock sector $\mathcal{H}_\chi$ (just as in \eqref{Hilbert}) and the \textit{total} GFT Hilbert space given by the tensor product of single-mode Hilbert spaces $\mathcal{H}_{\varphi_J}$ for the various modes.

Even though the Hamiltonian $\hat{{H}}_\varphi^\text{tot}$ (obtained by quantising \eqref{Htot}) couples modes in pairs ($J$ is coupled to $-J$), we show in appendix \ref{AppA} that it can be written as the sum over modes of single-mode contributions. As mentioned in section \ref{GFTSec}, one can distinguish between two cases based on the relative signs of $K^{(0)}_J$ and $K^{(2)}_J$, which lead to harmonic oscillator (HO) and squeezing (SQ) Hamiltonians. Then, defining $\mathfrak{J}^{\text{HO}}$ as the set of $J$ such that $K^{(0)}_J$ and $K^{(2)}_J$ have the same sign and $\mathfrak{J}^{\text{SQ}}$ as the set of $J$ such that $K^{(0)}_J$ and $K^{(2)}_J$ have opposite signs,\footnote{We exclude the fine-tuned case in which $K^{(0)}_J=0$ (note that the Hamiltonian \eqref{Htot} is not defined when $K^{(2)}_J=0$).} one can write the quantised version of \eqref{Htot} as
\begin{equation}\label{sumsingle}
	\hat{{H}}_\varphi^\text{tot}  = \sum_{J \in \mathfrak{J}^{\text{HO}}} \hat{H}_J^{\text{HO}} + \sum_{J \in \mathfrak{J}^{\text{SQ}}} \hat{H}_J^{\text{SQ}} 
\end{equation}
where, depending on the set $J$ belongs to,
\begin{equation}\label{2hams}
	\begin{aligned}
		\hat{H}_J^{\text{HO}} &= \epsilon_J \omega_J \left(\hat{a}^\dagger_J \hat{a}_J + \frac{1}{2}\right) \,,   &&J\in   \mathfrak{J}^{\text{HO}} \,,\\
		\hat{H}_J^{\text{SQ}}&= \frac{1}{2}\omega_J \left( (\hat{a}^\dagger_J)^2  + \hat{a}^2_J \right)\,, && J \in \mathfrak{J}^{\text{SQ}} \,,
	\end{aligned}
\end{equation}
with $\epsilon_J = (-1)^{\sum_I (j_I-m_I)}$ and $\omega_J = -{\rm sgn}\big( K_J^{(0)}\big) \sqrt{| K^{(0)}_J/K^{(2)}_J|}$. In \eqref{2hams} we are suppressing the explicit mentioning of the identity operators acting on all the other factors of $\mathcal{H}^\text{tot}_\varphi$; i.e., a tensor product with $\bigotimes_{J'\neq J} \mathbb{I}_{\varphi_{J'}}$ is understood for every single-mode Hamiltonian. Both single-mode Hamiltonians \eqref{2hams} admit a spectral decomposition (we refer to appendix \ref{AppA} for details), so that the total Hamiltonian can be written as
\begin{equation}\label{superdecomp}
	\hat{{H}}_\varphi^\text{tot} = \sum_J
	\left({{\intsum}_E} E_J |\psi^{E_J}\rangle \langle \psi^{E_J} | \right) \,,
\end{equation}
where the sum-integral notation introduced in \cite{Trinity,RelativisticTrinity} is used to take into account all modes in a compact way (it represents a sum for modes $J \in \mathfrak{J}^{\text{HO}}$ and an integral $\int_\pm$ for modes $J \in \mathfrak{J}^{\text{SQ}}$). Accordingly, the notation $|\psi^{E_J}\rangle$ refers to harmonic oscillator number eigenstates associated with \textit{discrete} energy eigenvalues $E_J$ for $J \in \mathfrak{J}^{\text{HO}}  $,\footnote{Specifically, $E_J = \epsilon_J \omega_J \left(n + \frac{1}{2}\right)$ for $J \in \mathfrak{J}^{\text{HO}}  $, with $n \in \mathbb{N}_0$.} and squeezing Hamiltonian eigenstates $|\psi^{E_J}_\pm\rangle$ introduced in section \ref{QuantumSec} (see also appendix \ref{AppA}), with \textit{continuous} label $E_J$ and degeneracy label $\pm$, for $J \in \mathfrak{J}^{\text{SQ}}$.

Since every single-mode Hamiltonian only acts on the corresponding $\mathcal{H}_{\varphi_J}$ (namely the factor of \eqref{bigHilbert} with the same $J$), one can easily show that
\begin{equation}
	\hat{{H}}_\varphi^\text{tot} \left( \bigotimes_J
	|\psi^{E_J} \rangle \right) = E_\text{tot} \left( \bigotimes_J
	|\psi^{E_J} \rangle \right) \,,
\end{equation}
where $\bigotimes_J 
|\psi^{E_J}\rangle  \in \mathcal{H}^{\text{tot}}_\varphi$ and $E_\text{tot}:=\sum_J
E_J$. Then, one can use the quantum constraint, $\hat{C} =  \hat{p}_\chi\otimes \mathbb{I}_\varphi^\text{tot} + \mathbb{I}_\chi \otimes \hat{H}^\text{tot}_\varphi$ with $\mathbb{I}_\varphi^\text{tot} = \bigotimes_J   \mathbb{I}_{\varphi_J}$, to formally define a group averaging operation $\delta(\hat{C})$ (see \eqref{deltaC}) to obtain physical states. As in the single-mode scenario, one starts with a kinematical state
\begin{equation}\label{kinstate5a}
	|\Psi_{\text{kin}}\rangle =  {{\intsum}_{\{E_J\}}}  \int \dd p_\chi \, \Psi \left(p_\chi, \{E_J\}, \{\pm\} \right) \, |p_\chi\rangle \otimes \left(\bigotimes_J 
	|\psi ^{E_J}\rangle\right)\,,
\end{equation}
where $\Psi (p_\chi, \{E_J\}, \{\pm\})$ depends on discrete $E_J$ variables for all $J \in \mathfrak{J}^{\text{HO}}$ and on continuous $E_J$ variables {(together with the set of all $\pm$ labels)} for all $J \in \mathfrak{J}^{\text{SQ}}$, which are respectively all summed and integrated over with the notation ${{\intsum}_{\{E_J\}}}$. Then, a physical state $|\Psi_{\text{phys}} \rangle= \delta (\hat{C}) |\Psi_{\text{kin}}\rangle$ reads
\begin{equation}\label{physmm}
	|\Psi_{\text{phys}}\rangle = {{\intsum}_{\{E_J\}}} \Psi \left(-E_\text{tot}, {\{E_J\}} , \{\pm\} \right)\, |-E_\text{tot}\rangle \otimes \left(\bigotimes_J
	|\psi ^{E_J}\rangle\right) \,.
\end{equation}

By suitably generalising the construction for multiple field modes, the Page--Wootters formalism described in section \ref{QuantumSec} can be applied here. In particular, one can define the conditioned state $	|\psi(\chi_0)\rangle_{\varphi_{\text{tot}}} \in  \mathcal{H}^{\text{tot}}_\varphi$ as
\begin{equation}\label{multiconditioned}
	|\psi(\chi_0)\rangle_{\varphi_{\text{tot}}} : = \Big(\langle \chi_0| \otimes \mathbb{I}_\varphi^\text{tot}\Big) |\Psi_{\text{phys}}\rangle \,,
\end{equation}
and check that both the Page--Wootters inner product (defined as in \eqref{PWIP} with $\mathbb{I}_\varphi$ replaced by $\mathbb{I}_\varphi^\text{tot}$) and the physical inner product \eqref{PIP} evaluate to
\begin{equation}\label{IPmm}
	\langle \Psi_{\text{phys}}|\Psi_{\text{phys}}\rangle_{\text{PW}}  = \langle \Psi_{\text{phys}}|	\Psi_{\text{phys}}\rangle_{\text{phys}} = {{\intsum}_{\{E_J\}}} \big|\Psi \left(-E_\text{tot}, {\{E_J\}}, \{\pm\} \right)\big|^2 \,.
\end{equation}

Since the procedure is the same as in the single-mode case, we can readily show the main results as follows. First, one defines the quantum version of \eqref{multiRDO} for the total number operator $\hat{\mathfrak{N}}_{\text{tot}} := \sum_J
[\hat{\mathfrak{a}}_J^\dagger\hat{\mathfrak{a}}_J \otimes (\bigotimes_{J'\neq J}\mathbb{I}_{\varphi_{J'}})]$ generalising \eqref{DiracN} as
\begin{equation}\label{totalND}
	\hat{N}_D^{\text{tot}}(\chi_0)
	: =  \frac{1}{2 \pi} \int \dd s \, \Big[\hat{U}_\chi (s) \otimes \Big(\bigotimes_J
	\hat{U}_{\varphi_J} (s)\Big)\Big] \left( |\chi_0\rangle \langle \chi_0 | \otimes \hat{\mathfrak{N}}_{\text{tot}}   \right)\Big[\hat{U}^\dagger_\chi (s) \otimes \Big(\bigotimes_J
	\hat{U}^\dagger_{\varphi_J} (s)\Big)\Big] \,,
\end{equation}
where the tensor product of operators $\hat{U}_\chi (s) \otimes \left(\bigotimes_J
\hat{U}_{\varphi_J} (s)\right) := e^{-{\rm i} s\hat{p}_\chi } \otimes \left(\bigotimes_J 
e^{-{\rm i} s \hat{H}_J}\right)$ leaves the physical state \eqref{physmm} invariant (cf.\ \eqref{timeless}). Then, exactly as in \eqref{equivalenceN} and \eqref{result}, one shows that
\begin{equation}\label{multimoderesult}
	N_D^{\text{tot}}(\chi_0) := 	\langle \Psi_{\text{phys}} | \hat{N}_D^{\text{tot}}(\chi_0) |\Psi_{\text{phys}}\rangle_{\text{phys}} 
	=  	\langle \Psi_{\text{phys}} | \hat{\mathfrak{N}}_{\text{tot}} |\Psi_{\text{phys}}\rangle_{\text{PW}} = {}_{\varphi_{\text{tot}}} \langle \psi(\chi_0) |  \hat{\mathfrak{N}}_{\text{tot}}   | \psi(\chi_0) \rangle_{\varphi_{\text{tot}}} \,,
\end{equation}
where we used the conditioned state of the Page--Wootters formalism \eqref{multiconditioned} in the last equality. Recalling that the conditioned state satisfies a Schrödinger equation with respect to $\chi_0$ (cf.\ \eqref{Schro}), we recover the results from the deparametrised approach for a GFT with multiple modes since, working in the Heisenberg picture, \eqref{multimoderesult} is
\begin{equation}\label{sumadaga}
	\begin{aligned}
		N_D^{\text{tot}}(\chi_0) 
		& = {}_{\varphi_{\text{tot}}}\langle \psi |	\Big[\bigotimes_J
		\hat{U}^\dagger_{\varphi_J} (\chi_0)\Big] \hat{\mathfrak{N}}_{\text{tot}} \Big[\bigotimes_J
		\hat{U}_{\varphi_J} (\chi_0)\Big] |\psi \rangle_{\varphi_{\text{tot}}} \\
		& =   		{}_{\varphi_{\text{tot}}}\langle \psi | \, \sum_J
		\hat{U}^\dagger_{\varphi_J} (\chi_0) \, \hat{\mathfrak{a}}^\dagger_J \hat{\mathfrak{a}}_J \, \hat{U}_{\varphi_J} (\chi_0)  \, | \psi \rangle_{\varphi_{\text{tot}}} \,,
	\end{aligned}
\end{equation}
namely the expectation value of \eqref{finalresult} (again note that a tensor product with $\bigotimes_{J'\neq J} \mathbb{I}_{\varphi_{J'}}$ is understood in the last line of \eqref{sumadaga}). 

In summary, this section simply shows that one can properly introduce both a Dirac quantisation and the Page--Wootters formalism for a parametrised GFT with any number of Peter--Weyl modes, and obtain a generalisation of all the results of sections \ref{ClassicalSec} and \ref{QuantumSec}. From a conceptual point of view, everything is analogous to the single-mode case: the quantum theory obtained from our parametrised GFT is still characterised by a Hilbert space split into clock and geometry (cf.\ \eqref{bigHilbert}). By carefully adapting the formalism to accommodate multiple GFT modes, one finds a clear notion of physical relational observables representing quantum geometrical quantities (for instance, the total number of GFT quanta in the various modes \eqref{sumadaga}) for every value of relational time. 

While we have assumed a maximum value for the Peter--Weyl label $J_\text{max}$ in the above scenario, one might be able to remove the cut-off and consider the full GFT using the theory of \textit{infinite tensor products} firstly developed in \cite{vonNeumann} and implemented in the context of loop quantum gravity in \cite{Thiemann2000,Sahlmann2001}. In particular, extra care would be needed to make sure the inner product \eqref{IPmm}, which in principle can contain infinite sums and integrals, converges. Such questions require functional analysis techniques that are beyond the scope of the present paper, and we leave them for future work.

\subsection{Many reparametrisation symmetries and multi-fingered time}\label{MultimodeSecB}

Here we go beyond the simple generalisation of section \ref{MultimodeSecA}, looking into the possibility of having different clocks for different GFT modes. In the quantum theory of section \ref{QuantumSec}, which focuses on a single group field mode, the Hilbert space \eqref{Hilbert} seems to have an ``artificial'' symmetry as it splits into two equal pieces, one for the matter clock and one for the GFT mode. This is not the case for the theory of section \ref{MultimodeSecA}, where we consider multiple modes on the GFT sector, but we still make use of one single clock (cf.\ \eqref{bigHilbert}). In this section we restore such a symmetry so that we have a clock Hilbert space that is ``as big'' as the GFT Hilbert space. In other words, we study the case where every GFT mode $J$ has its own relational time, which realises what is known as \textit{multi-fingered time evolution} \cite{Kuchar_Bubble,Kuchar_generic}, as we explain below. This relates to some of the ideas of \cite{Marchetti2021} where the various GFT quanta are associated with different ``single-quantum times''; however, in that scenario one encounters the problem of synchronisation since it is not clear how to find a unique time variable to describe the whole many-body system (see also the earlier work \cite{IshaDaniele} for a classical picture). Here, we will show that by working with field modes -- and ``single-mode times'' -- rather than particles, one can still obtain well-defined Dirac observables and therefore pose meaningful relational dynamical questions.

In order to discuss ``multiple times'' and hence generalise the single reparametrisation invariance of section \ref{MultimodeSecA}, we employ the following trick. First, we rewrite the action \eqref{DepAction} as a sum over single-$J$ contributions, $S_0[\varphi] =  \int \dd \chi \sum_J \mathcal{L}_J$ (see the end of appendix \ref{AppA} for the details); that is, we consider the action \eqref{sumL}. We then move the integral sign under the summation sign and rename dummy integration variables such that, for every element in the sum over $J$, $\chi$ gets labelled with an index as $\chi_J$. This allows to write the action equivalently as
\begin{equation}
	S_0 [\varphi] = \frac{1}{2} \sum_J \int \dd \chi_J    \left(K_J^{(0)}\varphi^2_J(\chi_J) -  K_J^{(2)} \big({\partial_{\chi_J}}\varphi_J(\chi_J)\big) ^2\right) \,.
\end{equation}
Then, one can apply the parametrisation strategy adopted in previous sections, but here for every mode $J$. In other words, one can parametrise the theory multiple times (adding a symmetry for every $J$) by introducing a set of parameters $\tau_J$. Just as in section \ref{ClassicalSec}, this process essentially doubles the phase space and one can write the parametrised action as
\begin{equation}\label{SB}
	\begin{aligned}
		S [\varphi,\chi] &=  \frac{1}{2} \sum_J	\int  \dd \tau_J \, \left( K^{(0)}_J\varphi^2_J \, \partial_{\tau_J} \chi_J  -  K_J^{(2)}\frac{(\partial_{\tau_J}\varphi_J)^2}{\partial_{\tau_J} \chi_J } \right) \\&= \frac{1}{2}\int  \dd \tau \, \sum_J	\left( K^{(0)}_J\varphi^2_J \, \partial_\tau \chi_J  -  K_J^{(2)}\frac{(\partial_\tau\varphi_J)^2}{\partial_\tau \chi_J } \right)
		\,.
	\end{aligned}
\end{equation}
The first line of \eqref{SB} explicitly shows multiple reparametrisation invariances (one for every mode) and in the second line we renamed the dummy integration variables $\tau_J$ to $\tau$, so that we could write the action as a single integral. Note that while in the action \eqref{supar} the same $\chi$ contributes to all the terms in the sum, in \eqref{SB} we have a different $\chi_J$ for every $J$. With the usual steps, one can easily derive the Hamiltonian theory from \eqref{SB}. The conjugate momenta to $\varphi_J$ and $\chi_J$ are defined for every mode $J$ as
\begin{equation}
	\pi_J = -K^{(2)}_J \frac{\partial_\tau \varphi_{J}}{\partial_\tau \chi_J} \,, \qquad \qquad 
	p_{\chi_J}  = \frac{1}{2} \left( K^{(0)}_J\varphi^2_J +K^{(2)}_J \frac{ (\partial_\tau \varphi_J)^2}{(\partial_\tau \chi_J)^2}  \right)  \,.
\end{equation}
Moreover, the theory is subject to a set of \textit{independent} first-class constraints (one for every mode),
\begin{equation}\label{CJ}
	C_J = H_J + p_{\chi_J} = 0 \,, \qquad \qquad 	H_J = -\frac{1}{2} \left[\frac{\pi^2_{J}(\chi)}{K^{(2)}_{J}}+K^{(0)}_{J}\varphi^2_{J}(\chi)\right] \,.
\end{equation}
While the constraint \eqref{CA} was associated with a single lapse function, one here has a mode-dependent lapse $N_J$ for every constraint \eqref{CJ}. Indeed, just like with \eqref{S1} and \eqref{Saa}, it is easy to see that this theory can be obtained from an action
\begin{equation}\label{SMF}
	S[\varphi, \chi] = \int \dd \tau \, \sum_J  \left(
	\pi_J \partial_\tau \varphi_{J} +p_{\chi_J} \partial_\tau \chi_J - N_J C_J \right) \,,
\end{equation}
which explicitly shows that the total super-Hamiltonian is given by $\sum_J N_J C_J$. Since the modes are independent, one obtains the equations of motion
\begin{equation}\label{lapses}
	\partial_\tau \varphi_J = \{ \varphi_J ,{\textstyle \sum_J} N_J C_J \}= - N_J \frac{\pi_{J}}{K^{(2)}_J} \,, \qquad \quad
	\partial_\tau \chi_J =\{ \chi_J  ,{\textstyle\sum_J} N_J C_J \}= N_J \,,
\end{equation}
clearly showing that the lapses $N_J$ describe the rate of change of the various $\chi_J$ (which will be used as clocks) with respect to $\tau$.

Following \cite{DittrichDO,GieselDO,Tambornino,ThiemannCanAlsoDoDiracObs}, one can generalise the notion of relational Dirac observables to systems with multiple constraints. In particular, the scenario discussed here is easily tractable since the constraints form an Abelian algebra,\footnote{Not many physical systems exhibit this property. An example with multiple constraints forming an Abelian algebra can be found in \cite{Campiglia_loopspherical,Gambini_LoopBH}, where loop quantum gravity techniques are applied to spherically symmetric settings.} $\{C_J,C_{J'}\}=0$. Because a gauge-orbit is multidimensional, one needs to introduce as many dynamical clocks as there are constraints. Then, similarly to the single-clock case, a complete observable can be defined as a relation between a phase space function $f_\varphi$ (of the $\varphi$ degrees of freedom only) and a set of independent clocks $\chi_J$. In short, the relational Dirac observable (generalising \eqref{classicalDO} and \eqref{multiRDO}) defined as \cite{DittrichDO,GieselDO,Tambornino,ThiemannCanAlsoDoDiracObs}
\begin{equation}\label{multifingeredDO}
	F_{f_\varphi,\{\chi_J\}}(\{\chi_J^0\}) := \sum_{n=0}^\infty \frac{1}{n!} \Big\{f_\varphi, \sum_{J} \alpha_J H_{J} \Big\}_n \Bigg|_ {\alpha_J \rightarrow (\chi_{J}^0 - \chi_{J}) } 
\end{equation}
is invariant under the flows generated by the constraints \eqref{CJ}. $F_{f_\varphi,\{\chi_J\}}(\{\chi_J^0\})$ gives the value of $f_\varphi$ when the dynamical clocks $\chi_J$ take the values $\chi_J^0$ for all $J$'s (the curly brackets in \eqref{multifingeredDO} are meant to emphasise that we deal with a set of clocks and not with the $J$-th clock only). The linear combination of the single-mode GFT Hamiltonians $\sum_{J} \alpha_J H_{J}$ can be used to define a physical Hamiltonian\footnote{Note that incorporating the $J$-dependent flow parameters into the definition of the Hamiltonian is equivalent to a rescaling of $\tau$ such that the initial and final configurations are separated by a time interval of length unity.} which generates evolution for the complete observables \eqref{multifingeredDO}. Indeed, since one can show that $\partial_{\chi_{J}^0} 	F_{f_\varphi,\{\chi_J\}}(\{\chi_J^0\}) = \{F_{f_\varphi,\{\chi_J\}}(\{\chi_J^0\}), {H}_J\} $, the quantity $\sum_{J} \alpha_J H_{J}$ is the generator of multi-fingered time evolution; it evolves observables along the various arbitrary parameters $\alpha_J$ (related to the lapse functions in \eqref{lapses}) associated with the clocks $\chi_J$ (see \cite{DittrichDO,GieselDO,Tambornino,ThiemannCanAlsoDoDiracObs} for details).

All the properties of the quantum theory corresponding to \eqref{SMF} easily follow by suitably generalising the constructions of previous sections. The canonical operators now satisfy $	\left[\hat{\varphi}_J, \hat{\pi}_{J'}\right] = {\rm i} \delta_{JJ'  }$ and $\left[\hat{\chi} _J , \hat{p}_{\chi _{J'}}\right] = {\rm i} \delta_{JJ'  } $, and the kinematical Hilbert space is defined as
\begin{equation}\label{hugehilbert}
	\mathcal{H}_{\text{kin}} = \mathcal{H}^\text{tot}_\chi \otimes \mathcal{H}^\text{tot}_\varphi = \Big(\bigotimes_J \mathcal{H}_{\chi_J}\Big) \otimes  \Big(\bigotimes_J	\mathcal{H}_{\varphi_J} \Big)
	= \bigotimes_J  \Big( \mathcal{H}_{\chi_J} \otimes \mathcal{H}_{\varphi_J} \Big)\,,
\end{equation}
where in the last equality we rearranged the single-mode Hilbert spaces to show that this theory can be seen as a tensor product over modes of the theory studied in section \ref{QuantumSec}. While we already dealt with a theory on $\mathcal{H}^\text{tot}_\varphi$ with multiple GFT modes in section \ref{MultimodeSecA}, it also follows from the structure in \eqref{hugehilbert} that all the kinematical considerations for the ``matter sector'' (cf.\ \eqref{chiproperties}--\eqref{lastchiprop}) described in section \ref{QuantumSec} apply here. Thus, all single-mode quantities described there will simply obtain a $J$ label here (for both the $\varphi$ and $\chi$ sectors). 

In order to follow Dirac's quantisation programme, we write down the quantum constraints (corresponding to \eqref{CJ}) on the Hilbert space \eqref{hugehilbert} as
\begin{equation}
	\hat{C}_J =  \hat{p}_{\chi_J} \otimes \big(\bigotimes_{J'\neq J} \mathbb{I}_{\chi_{J'}}\big) \otimes \mathbb{I}^\text{tot}_\varphi + \mathbb{I}_\chi^\text{tot} \otimes \hat{H}_{\varphi_J} \otimes \big(\bigotimes_{J'\neq J} \mathbb{I}_{\varphi_{J'}} \big) \,,
\end{equation}
where the various identity operators clarify that each $\hat{C}_J$ acts non-trivially only on the respective $J$-th piece of $\mathcal{H}^\text{tot}_\chi$ and $ \mathcal{H}^\text{tot}_\varphi$ in \eqref{hugehilbert}. A physical state is annihilated by all the constraints separately,
\begin{equation}
	\hat{C}_J |\Psi_\text{phys}\rangle =0 \qquad \qquad \forall\, J \,,
\end{equation}
and is defined via group averaging as $|\Psi_\text{phys}\rangle := \prod_J \delta (\hat{C}_J)  |\Psi_\text{kin} \rangle$, where a generic state $|\Psi_\text{kin} \rangle$ on \eqref{hugehilbert} reads
\begin{equation}\label{hugekin}
	|\Psi_\text{kin} \rangle = {{\intsum}_{\{E_J\}}}  \int \prod_J \dd p_{\chi_J} \, \Psi \left(\{p_{\chi_J}\}, {\{E_J\}}, \{\pm\}\right) \, \Big[ \bigotimes_J \Big(|p_{\chi_J}\rangle \otimes |\psi ^{E_J}\rangle\Big)\Big]\,.
\end{equation}
Here we adopt the notation of section \ref{MultimodeSecA} (see in particular \eqref{kinstate5a}), and $\int \prod_J \dd p_{\chi_J}$ means we integrate over the scalar field momenta for all modes. One obtains the physical states
\begin{equation}
	|\Psi_\text{phys}\rangle= {{\intsum}_{\{E_J\}}} \Psi(\{-E_J\},\{E_J\} , \{\pm\}) \Big[\bigotimes_J\Big(|-E_J\rangle \otimes |\psi^{E_J}\rangle\Big) \Big] \,,
\end{equation}
which naturally generalises all the findings of previous sections (cf.\ \eqref{phys1m} and \eqref{physmm}). Aiming to discuss again the equivalence between the physical inner product with multiple constraints (denoted with an $M$), which we define as
\begin{equation}\label{multiphysIP}
	\langle \Psi_\text{phys}|\Psi_\text{phys}\rangle_\text{phys}^\text{M} := \langle \Psi_\text{kin}| \prod_J \delta(\hat{C}_J) |\Psi_\text{kin}\rangle_\text{kin}\,,
\end{equation}
and the Page--Wootters inner product, one needs to extend the Page--Wootters construction to the case of multiple clocks. In short, by introducing the ``multi-conditioned'' state in $ \mathcal{H}^\text{tot}_\varphi$
\begin{equation}\label{multicond}
	|\psi(\{\chi_J^0\}) \rangle_{\varphi_\text{tot}}^\text{M}:=  \Big[  \Big(\bigotimes_J \langle \chi_J^0|\Big) \otimes \mathbb{I}_\varphi^\text{tot}\Big] |\Psi_{\text{phys}}\rangle \,,
\end{equation}
which generalises \eqref{conditioned} by ``projecting'' \textit{all} clocks to the values $\chi_J^0$,\footnote{Note that a physical state can be written as $|\Psi_\text{phys}\rangle = \Big(\bigotimes_J \int \dd \chi_J^0 |\chi_J^0\rangle\Big) \otimes |\psi(\{\chi_J^0\})\rangle_{\varphi_\text{tot}}^\text{M} $ so that it encodes information about all the timelines in their entirety, and can be dubbed ``{state of histories}'', generalising \eqref{history}.} one is lead to the following definition of the multi-fingered ($M$) time Page--Wootters inner product
\begin{equation}\label{multiPWIP}
	\langle \Psi _\text{phys}|\Psi_\text{phys} \rangle_{\text{PW}}^\text{M} : = 	\langle \Psi _\text{phys}| \Big[  \Big(\bigotimes_J |\chi_J^0\rangle \langle \chi_J^0|\Big) \otimes \mathbb{I}_\varphi^\text{tot}\Big]  |\Psi_\text{phys} \rangle_{\text{kin}}\,.
\end{equation} 
Then, as expected from the results for a single clock of \cite{Trinity,RelativisticTrinity}, one finds
\begin{equation}
	\langle \Psi_\text{phys}|\Psi_\text{phys}\rangle_\text{phys}^\text{M} = 	\langle \Psi _\text{phys}|\Psi_\text{phys} \rangle_{\text{PW}}^\text{M} = {{\intsum}_{\{E_J\}}} |\Psi \left(\{-E_J\},\{E_J\},  \{\pm\}\right)|^2\,.
\end{equation}
We point out that the steps presented here represent the first explicit construction of the Page--Wootters formalism extended to the case of multiple (but finitely many) clocks, which corresponds to a Dirac quantisation with multiple Hamiltonian constraints. We note however that this could be seen as a special case of \cite{Hoehn_PWFT}, where the Page--Wootters formalism is formally applied to field theories. As mentioned above, the limiting case of infinitely many modes (and thus clocks) could be related to infinite tensor products techniques, and indeed be dealt with wave-functional treatments for quantum field theories (which are employed in \cite{Hoehn_PWFT}).

We now proceed to the formal quantisation of the classical observable \eqref{multifingeredDO},
\begin{equation}\label{QmultifingeredDO}
	\begin{aligned}
		\hat{F}_{f_\varphi, \{\chi_J\}}(\{\chi_J^0\})& :=\Big[\bigotimes_J \Big(\int \frac{\dd \chi_J}{2\pi } |\chi_J\rangle \langle \chi_J | \Big)\Big] \otimes \sum_n^\infty \frac{{\rm i}^n}{n!} \Big[\hat{f}_\varphi \, , \sum_{J} \alpha_J \hat{H}_{{J}}\big]_n \Bigg|_{\alpha_J \rightarrow(\chi_{J}-\chi_{J}^0)}\,,
	\end{aligned}
\end{equation}
which represents a natural generalisation of the quantity \eqref{Fquantum} introduced in \cite{Trinity,RelativisticTrinity}. In this sense, \eqref{QmultifingeredDO} extends the discussion of Dirac observables (and their connection to the Page--Wootters formalism) to the multi-fingered time scenario, which we now apply to our GFT setting.

From the physical Hamiltonian $\sum_{J} \alpha_J \hat{H}_{{J}}$ one can define the multi-fingered evolution operator $\bigotimes_J \hat{U}_{\varphi_J} (\alpha_J): = e^{-{\rm i} \sum_{J} \alpha_J\hat{H}_{{J}}}$ on $\mathcal{H}^\text{tot}_{\varphi}$, where the factors $\hat{U}_{\varphi_J} (\alpha_J)= e^{-{\rm i} \alpha_J \hat{H}_J}$ generalise the evolution operators of previous sections (cf.\ \eqref{Fquantumuseful}) by evolving along the mode-dependent time parameter $\alpha_J$. Then, specialising the expression \eqref{QmultifingeredDO} to the GFT number operator with $\hat{f}_\varphi = \hat{\mathfrak{N}}_\text{tot}$ (as in section \ref{MultimodeSecA}), one can define
\begin{equation}\label{NtotDmulticlock}
	\begin{aligned}
		\hat{N}^{\text{tot}}_{D}(\{\chi_J^0\}) & :=\Big[\bigotimes_J \Big(\int \frac{\dd \chi_J}{2\pi } |\chi_J\rangle \langle \chi_J | \Big)\Big] \otimes  \Big[ \Big(\bigotimes_J \hat{U}_{\varphi_J}(\chi_J-\chi_J^0)\Big) \hat{\mathfrak{N}}_\text{tot} \Big(\bigotimes_J \hat{U}^\dagger_{\varphi_J}(\chi_J-\chi_J^0)\Big)\Big]\\
		& = \int\prod_J \frac{\dd \alpha_J}{2\pi} \; \hat{U}_{\chi\varphi}^\text{tot}(\{\alpha_J\})\Big[\Big(\bigotimes_J 	|\chi_J^0 \rangle \langle \chi_J^0 | \Big)  \otimes \hat{\mathfrak{N}}_\text{tot} \Big]\hat{U}^{\text{tot}\, \dagger}_{\chi\varphi}(\{\alpha_J\}) \,,
	\end{aligned}
\end{equation}
where $\hat{U}_{\chi\varphi}^\text{tot}(\{\alpha_J\}) :=\bigotimes_J\Big(\hat{U}_{\chi_J}(\alpha_J)\otimes \hat{U}_{\varphi_J}(\alpha_J)\Big)$ is the generalisation of \eqref{Uchiphi}, we used the properties of the clock states introduced in section \ref{QuantumSec} (which now apply to all the clocks $\chi_J$), and we changed integration variables $\chi_J\rightarrow\alpha_J+\chi_J^0$ in the last line. We emphasise that $\hat{N}^{\text{tot}}_{D}(\{\chi_J^0\})$ is different from \eqref{totalND} as it depends on multiple clocks and thus it represents a multi-fingered time version of the number operator. Recall that the subscript $D$ means that \eqref{NtotDmulticlock} defines a true Dirac observable, which strongly commutes with \textit{all} the constraints, namely
\begin{equation}\label{FmC}
	\left[ \hat{N}^{\text{tot}}_{D}(\{\chi_J^0\}) , \hat{C}_J \right] =0  \qquad \qquad \forall\, J \,.
\end{equation}

Finally, since $\hat{N}^{\text{tot}}_{D}(\{\chi_J^0\})|\Psi_\text{phys}\rangle= \prod_J \delta(\hat{C}_J) \big[\big(\bigotimes_J |\chi_J^0\rangle \langle \chi_J^0 |\big) \otimes\hat{\mathfrak{N}}_\text{tot} \big] |\Psi_\text{phys}\rangle$, one can use the inner products \eqref{multiphysIP} and \eqref{multiPWIP}, as well as the state \eqref{multicond}, to show that
\begin{equation}
	\begin{aligned}\label{mfresult}
		\langle \Psi_\text{phys}|\hat{N}^{\text{tot}}_{D}(\{\chi_J^0\}) |\Psi_\text{phys}\rangle^\text{M}_\text{phys} & = \langle \Psi_\text{phys}| \hat{\mathfrak{N}}_\text{tot} |\Psi_\text{phys}\rangle^\text{M}_\text{PW}\\&= {}_{\varphi_{\text{tot}}}^{\hspace{2.5mm}\text{M}}\langle \psi(\{\chi_J^0\}) | \hat{\mathfrak{N}}_\text{tot}  | \psi(\{\chi_J^0\})\rangle_{\varphi_\text{tot}}^\text{M}\\& = 	
		{}_{\varphi_\text{tot}}^{\hspace{2.5mm}\text{M}}\langle \psi | \, \sum_J
		\hat{U}^\dagger_{\varphi_J} (\chi_J^0) \, \hat{\mathfrak{a}}^\dagger_J \hat{\mathfrak{a}}_J \, \hat{U}_{\varphi_J} (\chi_J^0) \, | \psi  \rangle_{\varphi_{\text{tot}}}^\text{M} \,,
	\end{aligned}
\end{equation}
where $|\psi\rangle_{\varphi_{\text{tot}}}^\text{M} = |\psi(0,\dots,0)\rangle_{\varphi_{\text{tot}}}^\text{M} $ and we suppressed again the trivial factors $\bigotimes_{J'\neq J}\mathbb{I}_{\varphi_{J'}}$. As expected, the last line in \eqref{mfresult} shows that every GFT mode in the total number operator evolves according to its own relational time parameter. 

At this point we want to stress that since in a free GFT the various $J$ modes are decoupled, a multi-fingered setting still allows to have well-defined Dirac observables, both classically and at the quantum level (in the sense that they commute with the constraints \eqref{FmC}). In other words, synchronisation between the different $J$-times is not really an issue here: while one can indeed reduce to the synchronised case of section \ref{MultimodeSecA} by gauge-fixing all the $\chi_J$ to be the same $\chi$ (using the freedom coming from all the reparametrisation symmetries), this is not necessary in order to answer physically meaningful questions. Indeed, one can study the dynamics of the various modes of the number operator by means of mode-dependent clocks, and still be able to compute the total number operator combining all the information (i.e., summing all the separate contributions together). In a sense, it does not matter whether one uses a unique time or a different time for every mode (even with an infinite number of modes), as they are independent. 

On this note, we also emphasise that the synchronisation problem mentioned in the literature \cite{Marchetti2021} relies on having different clocks for different GFT quanta, rather than field modes. Specifically, the issue arises because the number of particles is not conserved under time evolution (indeed, this is how the GFT cosmology framework explains the expansion of the Universe \cite{GFTcosmoLONGpaper,Gielen_2016,Oriti_2016,BOriti_2017,relham_Wilson_Ewing_2019,relhamadd}). Even in standard quantum field theory the number of particles is not a well-defined quantity, and the Fourier mode decomposition (here given by the Peter--Weyl decomposition \eqref{peterw}) represents the most natural way of reducing the theory to quantum-mechanical systems. Of course, one then interprets field excitations as particles so that the single-mode times of \eqref{mfresult} describe the evolution of GFT quanta associated with the same $J$ (naively, building blocks of quantum geometry with the same ``shape''). 

{Finally}, since adding reparametrisation symmetries does not change the physical content of the theory, we note that one could have investigated equivalent questions directly in the deparametrised setting of section \ref{GFTSec}. While there one usually studies observables where every mode is associated with the same time label (cf.\ \eqref{Ndep} and \eqref{finalresult}), \eqref{mfresult} suggests that one could have defined an observable as a sum of single-mode contributions \textit{at different times}, 
\begin{equation}\label{MultiDep}
	\hat{\mathcal{O}}_\text{tot} (\{\chi_J\}) = \sum_J e^{{\rm i}\hat{H}_J \chi_J} \hat{\mathcal{O}}_J(0) \, e^{-{\rm i}\hat{H}_J\chi_J}\,,
\end{equation}
where every Peter--Weyl mode is associated with a different reading of the same clock. In the free theory, this is a well-defined observable for the deparametrised approach that was never investigated because of its somewhat unusual interpretation (where the modes are observed at different times), which in section \ref{MultimodeSecB} is revisited in terms of multi-fingered time evolution. When $\hat{\mathcal{O}}=\hat{N}$, the expectation value of \eqref{MultiDep} is nothing but \eqref{mfresult}; in this sense, the articulated construction of section \ref{MultimodeSecB} shows again the equivalence between deparametrisation and ``post-quantum time'' dynamics.

\section{Discussion}

The main result of this paper is the construction and quantisation of parametrised group field theory models for quantum gravity coupled to a massless scalar field (to be used as a clock), and a proper definition of the corresponding relational quantum dynamics. By virtue of the equivalence between the quantum dynamics defined by relational Dirac observables and the Page--Wootters formalism, we showed that using the quantum degree of freedom associated to the scalar field as dynamical clock yields the same results as choosing the scalar field as background time variable at the classical level (i.e., following what we call {deparametrised approach} in section \ref{GFTSec}).

Specifically, we first analysed the case of a free parametrised GFT restricted to a single field mode. Employing the strategy of ``evolving constants of motion'' \cite{Rovelli_Amodel, Rovelli_Anhypothesis,RovelliOBS,RovelliPO}, we could define classical physical observables (which Poisson-commute with the Hamiltonian constraint) as relational Dirac observables \cite{DittrichDO,GieselDO,Tambornino}. In particular, we focussed on the number operator in the quantum theory -- the main geometrical quantity in GFT cosmology -- as evolving with respect to the values of the matter scalar field. After discussing the quantum kinematics of the geometry sector and of the clock sector, where the matter scalar field (and hence, ``time'') is represented by a quantum operator, we defined relational dynamics following both Dirac's method for constraint quantisation and the Page--Wootters formalism. These were shown to be equivalent in \cite{Trinity,RelativisticTrinity} for constraints of the form used in this paper. This is the first application of the Page--Wootters formalism in a non-perturbative quantum gravity theory, and it can provide useful insights for the interpretation of relational dynamics of quantum geometry. In particular, thanks to the projection of the physical state (annihilated by the Hamiltonian constraint) onto the ``conditioned state'' of the Page--Wootters formalism, this setup provides a coherent framework describing the evolution of (the expectation values of) the geometrical quantities of interest \textit{conditional} on a quantum time operator reading a certain value.

Other than reproducing the equivalence established in \cite{Trinity,RelativisticTrinity} for GFT, a central result of this paper is the recovery of the relational dynamics of the deparametrised setting. Before our work, it was not clear whether one was allowed to select a clock variable at the classical level without losing covariance \cite{Marchetti2021}. By matching the dynamics of the deparametrised approach to the ones of the ``tempus post quantum'' theory introduced here (which in particular has a clear notion of Dirac observables), we showed that choosing a clock and quantising are procedures that commute, at least for the free GFT models usually adopted to extract cosmology from quantum gravity.

We then generalised the construction to the more complicated case of free GFT models with an arbitrary (but finite) number of field modes. We distinguished between two scenarios: one where the system exhibits a single reparametrisation invariance (just like in the single-mode case) and one with many such invariances. The first case straightforwardly extends all the previously mentioned results; we showed that one can easily deal with observables for GFT models with as many mode-contributions as desired. In this sense, while the theory must be truncated to a finite number of modes for this construction to be well-defined, the level of approximation is arbitrary. 

The case of many reparametrisation symmetries, on the other hand, is achieved by parametrising the independent modes separately and relates to the idea of ``multi-fingered time'' introduced in \cite{Kuchar_Bubble,Kuchar_generic}. In addition to using Dirac's quantisation programme, which can take care of multiple constraints, we defined an extension of the Page--Wootters formalism to the case of multiple quantum clocks. This generalises and corroborates the correspondence between the two frameworks proved in \cite{Trinity,RelativisticTrinity} for a ``single notion of time''. We obtained a well-defined Dirac observable corresponding to the GFT number operator that describes each single-mode contribution as evolving with respect to a different ``single-mode time''. This theory represents the most general situation: one can (but does not need to) reduce to the setting with a single reparametrisation invariance thanks to the freedom -- coming from all the symmetries -- to ``synchronise'' the different mode-dependent notions of time, so as to have a unique clock for the entire system. Essentially, the framework introduced in this paper provides an appropriate representation at the quantum level for both cases (with one or multiple clocks for the various modes) as both can address well-defined questions regarding the relational dynamics of GFT observables, such as the number of quanta. 

We highlight that the ``problem of synchronisation'' noted in \cite{Marchetti2021} is related to the idea of GFT quanta (whose number is not conserved) as evolving with respect of different notions of time; this issue is avoided in our work because we defined multi-fingered time evolution with respect to field modes instead of particles. Already classically, the various modes can either evolve with respect to their own relational time or with respect to the same clock; we showed that for a non-interacting theory these dynamical questions are equally well-defined. Moreover, we note in passing that since our construction is built upon a well-defined physical inner product in the quantum theory, no divergences arise as in formalisms that aim to describe relational dynamics at the kinematical level.

In light of the results of \cite{Trinity,RelativisticTrinity} on the notion of changing temporal reference frames (see also \cite{Giacomini_qrf,Hohn_switch1,Hohn_switch2,Vanrietvelde_qrf}), our work opens up the possibility of investigating clock changes and obtaining a quantum notion of covariance in group field theory models.\footnote{Note that the role of diffeomorphisms in GFT has been studied by \cite{Baratin:2011tg}; these appear as global symmetries at the level of the group manifold (see also \cite{Gielen:2011dg}), and hence are not associated with constraints.} In particular, given that the GFT literature relies to a great extent on a single massless scalar field, one can now explore other clock candidates with a proper treatment at the quantum level. For example, building on the work of \cite{Axel_Steffen_scalars} for models with multiple scalar fields, one could take advantage of the ``clock-neutral'' picture to explicitly show the equivalence between quantum dynamics with different choices of clocks in GFT. Additionally, the tools developed here also allow to explore what happens if one uses a degree of freedom on the geometry sector (e.g., a GFT mode) as a clock to describe relational dynamics of the matter scalar field. In a sense, this reflects the situation in classical cosmology where for example the volume of an expanding Universe is expected to be a good clock. Potentially, these tools could be applied to other geometrical observables in GFT which may be considered as dynamical clocks (see \cite{Aniso} for a proposal with an anisotropy degree of freedom). All such investigations would provide a stronger handle on general covariance questions in a quantum gravity framework such as GFT; questions that might be of interest for the communities of quantum information and foundations of quantum mechanics, other than quantum gravity.

{As mentioned} in section \ref{MultimodeSec}, another natural line of research would be to study the case with infinite field modes, which requires functional analysis tools for the Page--Wootters formalism in the context of field theories \cite{Hoehn_PWFT}, and might be related to infinite tensor products techniques \cite{vonNeumann,Thiemann2000,Sahlmann2001}.

To conclude, we remark that this paper only deals with free group field theories; this is clearly a limitation as realistic models for quantum gravity would also include higher order terms in the group field. Importantly, such interactions are not expected to spoil the parametrisation process as one would still obtain a constraint that is linear in the scalar field momentum, and there would be no difference to the conceptual construction presented in this work. In particular, one could implement the Page--Wootters formalism and obtain a clock-neutral perspective; however, one would not be able to solve the interacting quantum theory explicitly as we do here. Finally, one could in principle also consider interactions between matter (i.e., the scalar field clock $\chi$) and geometry. Such interactions would require an explicit $\chi$ dependence in the action, which is usually excluded in GFT models for cosmology as $\chi$ is interpreted as a free massless scalar field (meaning the action must satisfy a shift symmetry). In this sense, a clock-system interaction is absent due to the very definition of the models, where the non-interacting massless scalar field is chosen because it behaves as an idealised clock. If one were to relax this assumption, the matter-geometry interactions could potentially be addressed with the general tools of \cite{SmithPW}, where the Page--Wootters states satisfy a time-nonlocal Schr\"odinger equation which can be solved perturbatively. We leave the investigation of models with possible nontrivial matter interaction for future work.

\begin{acknowledgments}
	This work was supported by the Royal Society through the University Research Fellowship Renewal URF$\backslash$R$\backslash$221005 (SG). 
\end{acknowledgments}

\appendix

\section{GFT Hamiltonian decomposition}\label{AppA}

In this appendix we show in detail how the total GFT Hamiltonian can be written as a sum of single-mode contributions, thus we obtain its spectral decomposition by studying properties of single-mode Hamiltonians. We begin by noting that the (quantum) Hamiltonian (cf.\ \eqref{relham}),
\begin{equation}\label{Hq}
	\hat{{H}}= -\frac{1}{2} \sum_{J} \left(\frac{\hat{\pi}_{J}(\chi)\hat{\pi}_{-J}(\chi)}{K^{(2)}_{J}}+K^{(0)}_{J}\hat{\varphi}_{J}(\chi)\hat{\varphi}_{-J}(\chi)\right)  \,,
\end{equation}
can be seen as the sum of \textit{two-mode} Hamiltonians, since every contribution to the sum depends on the modes $J$ and $-J$. In other words, even in the free theory, the Peter--Weyl modes are \textit{a priori} pairwise coupled such that the dynamics of the mode $J$ is coupled with that of the mode $-J$. Since the term within brackets in \eqref{Hq} is symmetric with respect to exchanging $J$ and $-J$ (recall that $K_J=K_{-J}$ by assumption), the sum receives the same contribution twice (except in the case where $J=-J$, i.e., a mode with vanishing magnetic indices). One thus obtains an overall factor of $2$, so that we are interested in summing over two-mode components of the form
\begin{equation}\label{contr}
	\hat{H}_{(J,-J)} = - \frac{\hat{\pi}_{J}(\chi)\hat{\pi}_{-J}(\chi)}{K^{(2)}_{J}}-K^{(0)}_{J}\hat{\varphi}_{J}(\chi)\hat{\varphi}_{-J}(\chi)\,.
\end{equation}
Excluding the specific modes for which $K^{(0)}_J = 0$, the contributions \eqref{contr} can be of two different types, depending on the signs of $K^{(0)}_{J}$ and $K^{(2)}_{J}$. The modes for which these have the same sign (we denote this case with a single dash $'$) contribute with a two-mode Hamiltonian $\hat{H}'_{(J,-J)}$ that automatically decouples the mode $J$ from $-J$ when working in the ladder operator basis. Indeed, using \eqref{aadag} one obtains
\begin{equation}\label{Hdash}
	\hat{H}'_{(J,-J)} = \epsilon_J \omega_J \left(\hat{a}_J^\dagger \hat{a}_J + \hat{a}^\dagger_{-J} \hat{a} _{-J} + 1 \right) = \hat{H}^{\text{HO}}_J + \hat{H}^\text{HO}_{-J} \,,
\end{equation}
with $\epsilon_J = (-1)^{\sum_I(j_I-m_I)} $ and $\omega_J = -{\rm sgn}\big( K_J^{(0)}\big) \sqrt{| K^{(0)}_J/K^{(2)}_J|}$. That is, $\hat{H}'_{(J,-J)}$ decouples into the sum of two single-mode harmonic oscillator (HO) Hamiltonians, which have well-known properties (eigenvalue problem, spectral decomposition, \textit{et cetera}). 

On the other hand, the modes for which $K^{(0)}_{J}$ and $K^{(2)}_{J}$ have opposite signs (we denote this case with a double dash $''$) contribute with a two-mode Hamiltonian $\hat{H}''_{(J,-J)}$ that takes the form
\begin{equation}\label{H''}
	\hat{H}''_{(J,-J)} =  \omega_J \left( \hat{a}^\dagger_J \hat{a}^\dagger_{-J} + \hat{a}_J \hat{a}_{-J}\right) \,.
\end{equation}
While this is still a two-mode operator coupling $J$ and $-J$, it is possible to diagonalise it as follows. First, we introduce the vector notation $\mathbf{a}^T := (\hat{a}_J , \hat{a}_{-J})$ and $(\mathbf{a}^\dagger)^T := (\hat{a}_J^\dagger , \hat{a}_{-J}^\dagger)$ which allows to rewrite \eqref{H''} as
\begin{equation}
	\hat{H}''_{(J,-J)} = \frac{1}{2} \omega_J \left( (\mathbf{a}^\dagger)^T \, {\sigma_1}\, (\mathbf{a}^\dagger) + (\mathbf{a})^T \, {\sigma_1}\, (\mathbf{a})	\right)\,,
\end{equation}
where here and in the following $\sigma_i$ refers to the $i$-th Pauli matrix. Then, we perform a change of basis by means of a unitary transformation $U$,
\begin{equation}\label{unitary}
	\mathbf{a} \; \longrightarrow \; U \mathbf{a} \,, \qquad \qquad  U := e^{- {\rm i} \frac{\pi}{4} \sigma_2}  \begin{pmatrix} 1 & 0 \\ 0 & -{\rm i} \end{pmatrix} \,.
\end{equation}
Finally, thanks to the property $e^{{\rm i}\frac{\pi}{4}\sigma_2} \, \sigma_1 \,	e^{{\rm -i}\frac{\pi}{4}\sigma_2} = \sigma_3$, one can easily check that $U^T \sigma_1 U = \mathbb{I}_{2\times2}$ so that $\hat{H}''_{(J,-J)}$ can be written in the diagonal form
\begin{equation}\label{Hddash}
	\hat{H}_{(J,-J)}'' 		 = \frac{1}{2} \omega_J \left( (\hat{a}_J^{\dagger})^2 + \hat{a}_J^2 \right) + \frac{1}{2} \omega_J \left( (\hat{a}_{-J}^{\dagger})^2 + \hat{a}_{-J}^2 \right) 
	= \hat{H}^{\text{SQ}}_J + \hat{H}^{\text{SQ}}_{-J} \,,
\end{equation}
namely as the sum of two single-mode squeezing (SQ) Hamiltonians. 

To summarise, all types of two-mode contributions $\hat{H}_{(J,-J)}'$ and $\hat{H}_{(J,-J)}''$ can be brought into a diagonal form (\eqref{Hdash} and \eqref{Hddash}); i.e., the Hamiltonians $\hat{H}_{(J,-J)}$ in \eqref{contr} decouple into a sum of single-mode Hamiltonians for all modes. It follows that the total GFT Hamiltonian can be cast as
\begin{equation}\label{sumsm}
	\hat{{H}}   = \sum_J \hat{H}_J \,,
\end{equation}
where, depending on the mode, $\hat{H}_J$ is either $\hat{H}^{\text{HO}}_J$ or $\hat{H}^{\text{SQ}}_J$ (as in the main text, see \eqref{sumsingle}).

To prove that one can obtain the spectral decomposition of the total Hamiltonian as given in \eqref{superdecomp}, one simply needs to work out the single-mode cases. Since the harmonic oscillator is solved in any quantum mechanics textbook, we briefly review here only the eigenvalue problem for the squeezing Hamiltonian. Working in the $(\hat{\varphi}, \hat{\pi})$ basis, the Hamiltonian $\hat{H}^{\text{SQ}}$ for a single mode (dropping the label $J$ in the following) can be written as (cf.\ \eqref{Ham})
\begin{equation}\label{IHOH}
	\hat{H}^\text{SQ} = \frac{1}{2} \sgn(K^{(0)}) \left( \frac{{\hat{\pi}}^2}{|K^{(2)}|} - |K^{(0)}| \hat{\varphi}^2 \right) \,.
\end{equation}
As pointed out in the main text, this is the Hamiltonian of an inverted (or upside-down) harmonic oscillator, for which we want to solve the Schr\"odinger problem \cite{Damped1,Damped2}. Given the shape of the potential, and in virtue of general properties for the one-dimensional Schr\"odinger equation (see, e.g., \cite{LandauQM}), the energy spectrum is continuous, $\sigma(\hat{H}^\text{SQ}) =(-\infty,\infty)$, and doubly-degenerate
\begin{equation}
	\hat{H}^\text{SQ}\psi^E_\pm(\varphi)= E \psi^E_\pm(\varphi)\,, \qquad \qquad E\in \mathbb{R} \,.
\end{equation}
Notice that the global sign factor $\sgn(K^{(0)})$ in \eqref{IHOH} is irrelevant since switching sign amounts to a relabelling of the eigenvalue $E \in (-\infty,\infty)$. Thus we can set $\sgn(K^{(0)}) =1$ for simplicity and without loss of generality in what follows. We then want to solve
\begin{equation} \label{solvethis}
	\frac{\dd^2}{\dd \varphi^2} \psi^E_\pm(\varphi) +\left(|K^{(0)}K^{(2)}| {\varphi}^2 +2 |K^{(2)}| E \right) \psi^E_\pm(\varphi) =0\,.
\end{equation}	
To that end, we introduce the variable 
\begin{equation}\label{zed}
	\zeta =\sqrt{2 {\rm i} \sqrt{|K^{(0)}K^{(2)}|}} \; \varphi 
\end{equation}
so that \eqref{solvethis} becomes
\begin{equation}\label{Weber}
	\frac{\dd^2}{\dd \zeta^2} \psi^E_\pm(\zeta) +\left( \nu + \frac{1}{2} -\frac{\zeta^2}{4} \right)\psi^E_\pm(\zeta) =0 \,,
\end{equation}
with
\begin{equation}\label{nu}
	\nu = - {\rm i}  \sqrt{\left|\frac{K^{(2)}}{K^{(0)}}\right|} \, E - \frac{1}{2} \,.
\end{equation}
\eqref{Weber} is called Weber equation and has known solutions in terms of parabolic cylinder functions \cite{abramowitzstegun,GradRy}, denoted $D_\nu(\zeta)$. Specifically, the two independent solutions are given by $\psi^E_+ (\zeta)= \mathcal{N}_+D_\nu(\zeta)$ and $\psi^E_-(\zeta)= \mathcal{N}_-D_{-\nu-1}({\rm i} \zeta)$, where $\mathcal{N}_\pm$ are normalisation constants and $\zeta$ and $\nu$ are given in \eqref{zed} and \eqref{nu}. Interestingly, when $\nu$ is a non-negative integer $n$, a parabolic cylinder function simplifies to $D_n(\zeta) = 2^{-n/2} e^{-\zeta^2/4} H_n(\zeta/\sqrt{2}) $, where $H_n$ is a Hermite polynomial (which notoriously solves the differential equation representing the eigenvalue problem for the standard harmonic oscillator). 

The first important property in the context of our paper is the orthonormality of the eigenstates $| \psi_\pm^{E} \rangle$ (switching to the bra-ket notation), which for suitable $\mathcal{N}_\pm$ reads \cite{Damped1,Damped2}
\begin{equation}\label{OrthonormalityIHO}
	\langle \psi_m^E| \psi_n^{E'} \rangle = \delta_{mn}\delta (E-E') \,,
\end{equation}
where $m$ and $n$ label the degeneracy ($+$ or $-$) and the Kronecker delta $\delta_{mn}$ indicates that ``$+$ states'' and ``$-$ states'' are orthogonal. \eqref{OrthonormalityIHO} is sometimes called generalised orthonormality because of the distributional nature of the Dirac delta (one can rigorously deal with distributions by considering rigged Hilbert spaces; see, e.g., \cite{ModernQM}). Moreover, one can decompose the Hamiltonian \eqref{IHOH} and obtain a spectral resolution in the following form \cite{Damped1,Damped2}
\begin{equation}\label{sqdecomp}
	\begin{aligned}
		\hat{H}^\text{SQ} &= \int \dd E \; E \; |\psi_+^E\rangle \langle \psi_+^E|+\int \dd E \; E \; |\psi_-^E\rangle \langle \psi_-^E| \\
		& = \int_\pm \dd E \; E \; |\psi_\pm^E\rangle \langle \psi_\pm^E|
		\,,
	\end{aligned}
\end{equation}
where we introduced the notation $\int_\pm :=\sum_\pm \int$. Finally, since the total GFT Hamiltonian \eqref{Hq} decomposes as \eqref{sumsm}, one can make use of \eqref{sqdecomp} and the standard properties of harmonic oscillator-like Hamiltonians to obtain a total spectral decomposition for $\hat{{H}}$ of the form given in \eqref{superdecomp}.

To conclude the appendix, we note that one could separate the theory into uncoupled modes already from classical considerations in the Lagrangian formalism, by writing the GFT action \eqref{DepAction} as a sum over single-mode contributions. In short, one can use the following classical field redefinitions that combine the $J$ and $-J$ modes (somewhat as in \eqref{unitary}),
\begin{equation}
	\varphi_J \; \longrightarrow \; \tilde{\varphi}_J = \frac{1}{\sqrt{2}} \left(\varphi_J +{\rm i} \varphi_{-J}\right) \,, \qquad \varphi_{-J} \; \longrightarrow \; \tilde{\varphi}_{-J} =\frac{1}{\sqrt{2}} \left(\varphi_J -{\rm i} \varphi_{-J}\right) \,,
\end{equation}
together with the corresponding ``velocities'' $\partial_\chi \tilde{\varphi}_J$ and $\partial_\chi \tilde{\varphi}_{-J}$, to show that the \textit{addenda} of the Lagrangian in \eqref{DepAction} can be written in the following form
\begin{equation}\label{LJLJ}
	\begin{aligned}
		\frac{1}{2}\left(	K_J^{(0)}\tilde{\varphi}_{-J}\tilde{\varphi}_J-  K_J^{(2)}\partial_\chi\tilde{\varphi}_{-J}\partial_\chi\tilde{\varphi}_J\right) 
		&=	 \frac{1}{4}  \left(K_J^{(0)}\varphi^2_J-  K_J^{(2)}(\partial_\chi\varphi_{J} )^2 + K_J^{(0)}\varphi^2_{-J}-  K_J^{(2)}(\partial_\chi\varphi_{-J} )^2\right)\\
		&=:\frac{1}{2}\left(\mathcal{L}_J + \mathcal{L}_{-J}\right)  \,.
	\end{aligned}
\end{equation}
Just as with the Hamiltonians (see discussion above), the Lagrangians $\mathcal{L}_J$ and $\mathcal{L}_{-J}$ provide the same contribution to the sum $\sum_J$ so that one can write the action \eqref{DepAction} as
\begin{equation}\label{sumL}
	S_0[\varphi] =\frac{1}{2} \int \dd \chi \sum_J \left(K_J^{(0)}\varphi^2_J-  K_J^{(2)}(\partial_\chi\varphi_{J} )^2\right) = \int \dd \chi \sum_J \mathcal{L}_J \,.
\end{equation}
The same conclusion was already reached in a different way in \cite{Axel_Steffen_scalars}.

%%%%%%%%%%%%%%%%%%%%%%%  Bibliography %%%%%%%%%%%%%%%%%%%%%%%
\let\oldaddcontentsline\addcontentsline
\renewcommand{\addcontentsline}[3]{}
\printbibliography
\let\addcontentsline\oldaddcontentsline

\end{document}